\documentclass{article}

\usepackage{arxiv}

\usepackage[utf8]{inputenc} 
\usepackage[T1]{fontenc}    
\usepackage{hyperref}       
\usepackage{url}            
\usepackage{booktabs}       
\usepackage{amsfonts}       
\usepackage{nicefrac}       
\usepackage{microtype}      
\usepackage{lipsum}

\usepackage{graphicx}          

\usepackage{amsfonts}       

\usepackage{amsmath,bm, bbm,amsthm}
\usepackage{xcolor}
\usepackage{comment}

\newtheorem{theorem}{Theorem}
\newtheorem{corollary}{Corollary}[theorem]
\newtheorem{lemma}[theorem]{Lemma}
\newtheorem{definition}{Definition}[theorem]
\newtheorem{proposition}{Proposition}[theorem]

\definecolor{ao(english)}{rgb}{0.0, 0.5, 0.0}
\definecolor{americanrose}{rgb}{1.0, 0.01, 0.24}
\definecolor{cerisepink}{rgb}{0.93, 0.23, 0.51}
\definecolor{darkorchid}{rgb}{0.6, 0.2, 0.8}
\definecolor{applegreen}{rgb}{0.55, 0.71, 0.0}
\definecolor{brightpink}{rgb}{1.0, 0.0, 0.5}
\definecolor{azure(colorwheel)}{rgb}{0.0, 0.5, 1.0}

\title{On the Equilibrium Elicitation of Markov Games Through Information Design}

\author{
  Tao Zhang\thanks{Corresponding author.} \\
  Department of Electrical and Computer Engineering\\
  New York University\\
  Brooklyn, NY, 11201\\
  \texttt{tz636@nyu.edu} \\
   \And
  Quanyan Zhu \\
  Department of Electrical and Computer Engineering\\
  New York University\\
  Brooklyn, NY, 11201\\
  \texttt{qz494@nyu.edu} \\
}

\begin{document}

\maketitle


\begin{abstract}

This work considers a novel information design problem and studies how the craft of payoff-relevant environmental signals solely can influence the behaviors of intelligent agents.
The agents' strategic interactions are captured by an incomplete-information Markov game, in which each agent first selects one environmental signal from multiple signal sources as additional payoff-relevant information and then takes an action.
There is a rational information designer (designer) who possesses one signal source and aims to control the equilibrium behaviors of the agents by designing the information structure of her signals sent to the agents.
An obedient principle is established which states that it is without loss of generality to focus on the direct information design when the information design incentivizes each agent to select the signal sent by the designer, such that the design process avoids the predictions of the agents' strategic selection behaviors.
We then introduce the design protocol given a goal of the designer referred to as obedient implementability (OIL) and characterize the OIL in a class of obedient perfect Bayesian Markov Nash equilibria (O-PBME).
A new framework for information design is proposed based on an approach of maximizing the optimal slack variables.
Finally, we formulate the designer's goal selection problem and characterize it in terms of information design by establishing a relationship between the O-PBME and the Bayesian Markov correlated equilibria, in which we build upon the revelation principle in classic information design in economics.
The proposed approach can be applied to elicit desired behaviors of multi-agent systems in competing as well as cooperating settings and be extended to heterogeneous stochastic games in the complete- and the incomplete-information environments.

\end{abstract}


\section{Introduction}


Building rational multi-agent system is an important research desideratum in Artificial Intelligence.
In goal-directed decision making systems, an agent's action is controlled by its consequence \cite{dickinson1985actions}.
In a game, the consequence of an agent's action is the outcome of the game, given as the reward of taking that action as well as the actions of his opponents, which situates the optimality criterion of each agent's decision making in the game.
A rational agent's reward may also depend on the payoff-relevant information, in addition to the actions.
The information may include the situation of the agents in a game, referred to as the state of the world, as well as his knowledge about his opponents' diverging interests and their preferences over the outcomes of the game.
Incorporating such payoff-relevant information in his decisions constitutes an essential part of an agent's rationality in the strategic interactions with his opponents.
Hence, one may re-direct the goal achievement of rational agents in a game by information provision.
In economics, this refers to as \textit{information design}, which studies how an information designer (she) can influence agents' optimal behaviors in a game to achieve her own objective, through the design of information provided to the game \cite{bergemann2019information}.

Referred to as the inverse game theory, mechanism design is a well-developed mathematical theory in economics that provides general principles of how to 
design \textit{rules of games} (e.g., rewarding systems with specifications of actions and outcomes) to influence the agents' strategic interactions and achieve system-wide goals while treating the information as given.
%
%
Information design, on the other hand, considers the circumstances when the information in the environment is under the control of the system designer and offers a new approach to indirectly elicit agents' behaviors by keeping the game rules fixed \cite{taneva2019information}.
%

This work considers a finite-agent infinite-horizon Markov game in an incomplete-information environment.
Each agent privately possesses a payoff-relevant information, called type, with a commonly known prior probability distribution.
At each period of time, agents observe a payoff-relevant global state (state).
In addition to the type and the state, each agent observes a batch of signals (signal batch, batch) at each period and then strategically chooses one signal from the batch as the additional information to his decision of actions.
Each agent's one-period reward is determined by his own action, the actions of his opponents, the global state, and his choice of signal.
We refer to this game as a \textit{base Markov game} (BMG).
The transition of the state, the prior of type, and the distribution of signals are referred to as the \textit{information structure} of the MBG.
%
%
In a BMG, each agent's behavior includes selecting a signal according to a \textit{selection rule} and taking an action according to a \textit{policy}.
Here, each agents' selection of signal and the choice of action are coupled since the selected signal enters the policy to determine the choice of the action.
If a mechanism designer aims to incentivize the agents to behave in her desired way, she directly modifies the BMG--\textit{reversing the game}--by changing the rules of encounters, including changing the reward function associated with actions and outcomes, while treating the information structure as  part of the environment.
%
%
An information designer, however, treats the BMG as fixed and modifies the information structure to elicit agents' equilibrium behaviors that coincide with her objective.

We study a novel dynamic information design problem in the BMG in which there are multiple sources of signals (signal sources, sources) and each of them sends one signal to each agent.
The signals sent by all sources constitute the signal batch observed by each agent at each time.
Among these sources, there is one rational information designer (referred to as \textit{principal}, she) who controls one signal source and intends to strategically craft the information structure of her signal by choosing a \textit{signaling rule} to indirectly control the equilibrium of the BMG.
We consider that other sources of signals provide additional information to the agents in a non-strategic take-it-or-leave-it manner.
%
%
The goal of the principal is to induce the agents to take actions according to an equilibrium policy that is desired by the principal.
However, the principal has no ability to directly program the agents' behaviors to force them to take certain actions.
Instead, her information design should provide incentive to rational agents to behave in her favor.
We study the extent to which the provision of signals along by controlling a single signal source can influence the agents' behavior in a BMG, when the agents have the freedom to choose any available signal in the batch.
We will name the BMG with a rational principal in this setting as an \textit{expanded Markov game} (EMG). 

Since the principal's design problem keeps the base game unchanged, our model fits the scenarios when the agents are intrinsically motivated and their internal reward systems translate information from external environment into internal reward signals \cite{chentanez2005intrinsically}. 
%
%
Intrinsically-motivated rational agents can be human decision makers with intrinsic psychological preferences or intelligent agents programmed with internal reward system.
The setting of multiple sources of additional information captures the circumstances when the environment is perturbed by noisy information, in which the agents may improperly use redundant and useless information to make their decisions that may deviate from the system designer's desire.
Also, the principal can be an adversary who aims to manipulate the strategic interactions in a multi-agent system through the provision of misinformation, without intruding each agent's local system to make any physical or digital modifications.


Although the principal's objective of information design in an EMG is to elicit an equilibrium policy, her design problem has to take into consideration how the agents select the signals from their signal batches because each agent's choice of action is coupled with his selection of signal. 
In an information design problem, the principal chooses an information structure such that each agent selects a signal using a selection rule and then takes an action according to a policy which matches the principal's goal.
We use \textit{admissibility} to denote the constraint such that the agents' equilibrium policy coincides with the principal's goal.
We characterize the information design problem in an EMG into two classes: the \textit{indirect} information design (IID) and the \textit{direct} information design (DID).
An IID is indirect in the sense that the signals sent by the principal may not be selected by some agents, thereby the actions taken by those agents are independent of the principal's (realized) signals. 
However, even though her signal does not enter an agent's policy to take an action, the principal can still influence the agent's action because the distribution of the signal batch is influenced by her information structure (given the distributions of signals from other sources) which affects the agents' selections of signals. Hence, the agents' behaviors indirectly depend on the principal's choice of information structure.
IID requires the principal to accurately predict each agents' strategic selection rules as well as the induced policies.
In DID problems, on the other hand, each agent always selects the signal sent by the principal and then takes an action.
Thus, the realizations of the principal's signals directly enter the agents' choice of actions.
In addition to the admissibility, another key restriction of the principal's DID problem is a notion of \textit{obedience} which requires that, with the information structure of the principal, each agent is incentivized to select the signal from the principal rather than choose one from other signal sources.
%
%
The key simplification provided by the DID is that the principal's prediction of the agents' strategic selection rules is replaced by a straightforward obedient selection rule that always prefers the principal's signals.

This paper makes three major contributions to the foundations of information design. 
First, we define a dynamic information design problem in an environment where the agents have the freedom to choose any available signal as additional information. An \textit{obedience principle} is established and formally states that for every IID that leads to an equilibrium policy, there exists a DID that leads to the same equilibrium policy.
As a result, the principal can focus on the DID of the EMG.
Captured by the notion of \textit{obedient implementability}, the principal's DID problem is constrained by the \textit{obedient} condition that incentivizes the agents to select the principal's signals and the \textit{admissibility} condition such that the agents take actions which meets the principal's goal equilibrium.
Our information design problem is distinguished from others in economics that study the commitment of the information design in a game when there is only a single source of additional information in static settings (e.g., \cite{mathevet2020information,taneva2019information,bergemann2016bayes,kamenica2011bayesian}) as well as in dynamic environment (e.g., \cite{ely2015suspense,passadore2015robust,doval2020sequential,ely2017beeps,ely2020moving,makris2018information}) and the settings in which the agents do not make a choice from multiple designers (e.g., \cite{koessler2018interactive}).
Second, we propose a new solution concept termed \textit{obedient perfect Bayesian Markov Nash equilibrium} (O-PBME), which allows us to handle the undesirable equilibrium deviations of agents in DID in a principled manner.
Specifically, by bridging our incomplete-information Markov game with dynamic programming and uncovering the close relationship between the O-MBNE and the optimization of the occupancy measures, we characterize the obedient implementability and explicitly construct the principal's DID problem.
Third, we formulate the principal's optimal goal selection problem and transform it to an optimal DID problem in which the admissibility condition is replaced by the optimality of the induced equilibrium policy with respect to the principal's objective.
A representation principle is obtained and formally states that the principal's goal selection from a set of equilibria referred to as the \textit{Bayesian Markov correlated equilibria} can be fully characterized by an information design that is implementable in an O-PBNE.

\subsection{Related Work}

We follow a growing line of research on creating incentives for interacting agents to behave in a desired way.
The most straightforward way is based on mechanism design approaches that properly provide reward incentives (e.g., contingent payments, penalty, supply of resources) by directly modifying the game itself to change the induced preferences of the agents over actions.
Mechanism design approaches have been fruitfully studied in both static \cite{myerson1981optimal} as well as dynamic environment \cite{pavan2014dynamic,zhang2019incentive,zhang2021differential}.
For example, auctions \cite{milgrom2004putting,bhat2019optimal} specify the way in which the agents can place their bid and clarify how the agents pay for the items; in matching markets \cite{sonmez2011matching,zhang2019optimal}, matching rules matches agents in one side of a market to agents of another side that directly affect the payoff of each matched individuals.
%
%
In reinforcement learning literature, reward engineering \cite{dewey2014reinforcement,nagpal2020reward,hadfield2017inverse} is similar to mechanism design that directly crafts the reward functions of the agents that post specifications of the learning goal.

Our work lies in another direction: the information design. 
Information design studies how to influence the outcomes of the decision makings by choosing signal (also referred to as signal structure, information structure, Blackwell experiment, or data-generating process) whose realizations are observed by the agents \cite{kamenica2019bayesian}.
In a seminar paper \cite{kamenica2011bayesian}, Kamenica and Gentzkow has introduced \textit{Bayesian persuasion} in which there is an informed sender and an uninformed receiver. 
%
The sender is endowed to commit to choosing any probability distribution (i.e., the information structure) of the signals as a function of the state of the world which is payoff-relevant to and unobserved by the receiver.
The Bayesian persuasion can be interpreted as a communication device that is used by the sender to inform the receiver through the signals that contain knowledge about the state of the world. Hence, the sender controls what the agent gets to know about the payoff-relevant state.
With the knowledge about the information structure, the receiver forms a posterior belief about the unobserved state based on the received signal. 
Hence, the information design of Bayesian persuasion is also referred to as an exercise in belief manipulation.
Other works alongside with the Bayesian persuasion include \cite{brocas2007influence,rayo2010optimal,arieli2019private,castiglioni2020online}.
In \cite{mathevet2020information}, Mathevet et al. extends the single-agent Bayesian persuasion of \cite{kamenica2011bayesian} to a multi-agent game and formulate the information design of influencing agents' behaviors through inducing distributions over agents' beliefs.
In \cite{bergemann2016bayes}, Bergemann and Morris have also considered information design in games. They have formulated the Myersonian approach for the information design in an incomplete-information environment. 
The essential of the Myersonian information design is the notion of Bayes correlated equilibrium, which characterizes the all possible Bayesian Nash equilibrium outcomes that could be induced by all available information structures.
The Myersonian approach avoids the modeling of belief hierarchies \cite{mertens1985formulation} and constructs the information design problem as a linear programming.
Information design has been applied in a variety of areas to study and improve real-world decision making protocols, including stress test in finance \cite{goldstein2018stress,inostroza2018persuasion}, law enforcement and security  \cite{hernandez2018bayesian,rabinovich2015information}, censorship \cite{gehlbach2014government}, routing system \cite{das2017reducing}, finance and insurance \cite{duffie2017benchmarks,szydlowski2021optimal,garcia2021information}.
Kamenica \cite{kamenica2019bayesian} has provided a recent survey of the literature of Bayesian persuasion and information design.

This work is based on Myersonian approaches and fundamentally differs from existing works on the information design.
First, we consider a different environment.
Specifically, we consider the setting when there are multiple sources of signals and each agent chooses one realized signal as an additional (payoff-relevant) information at each time.
Among these sources of signals, there is an information designer who controls one of these sources and aims to induce equilibrium outcomes of the incomplete-information Markov game by strategically crafting information structures.
Second, other than only taking actions, each agent in our model makes a coupled decision of selecting a realized signal and taking an action.
Hence, the characterization of the solution concepts in our work is different from the equilibrium analysis in other works.
Third, we also provide an approach with an explicit formulation to relaxing the optimal information design problem.

\textbf{Organization.}
The rest of the paper is organized as follows. Section \ref{sec:preliminary} describes the background and the basic concepts related to this work.
In Section \ref{sec:the_information_design_problem}, we describe the model and formally define the information design problem.
The notions of implementabilities are introduced to describe the optimality of the information design for the indirect and the direct settings.
The obedient perfect Bayesian Markov Nash equilibrium (O-PBME) is defined as the solution concept of our information design.
In Section \ref{sec:obedient_implementability} characterizes the obedient implementability in O-PBME by formulating an explicit design regime of the principle's information structure.
Section \ref{sec:conclusion} concludes the work.

\section{Preliminary: Finite-Player Game Model}\label{sec:preliminary}

\noindent \textbf{\textit{Convention.}} For the compactness of notations, we only show the elements, but not the sets, over which are summed under the summation operator. 
The notations are summarized in Appendix \ref{app:summary_notation}.

In this section, we review fundamental concepts in game theory to situate our contributions of this work.
%
This work focuses on games of $n$ self-interested agents, $N<\infty$, denoted by $\mathcal{N}\equiv[n]$, whose \textit{action space} is given as $\bm{\mathcal{A}}\equiv\{\mathcal{A}_{i}\}_{i\in\mathcal{N}}$. A typical agent is referred to as agent $i$, $i\in \mathcal{N}$.
\textit{Norm-form} (or strategic-form) is a basic representation of a static game:
\begin{definition}[Normal-Form Game \cite{ziebart2011maximum}] 
A normal-form game is defined by a tuple $\widehat{G}\equiv<\mathcal{N}, \bm{\mathcal{A}},  \bm{u}>$.
%
$\bm{u}\equiv\{u_{i}\}_{i\in \mathcal{N}}$, where $u_{i}: \bm{\mathcal{A}} \mapsto \mathbb{R}$ is the payoff function of agent $i\in \mathcal{N}$.
\end{definition}

A normal-form game considers that agents' payoffs for the outcomes of the game are common knowledge in equilibrium. 
Each agent $i\in \mathcal{N}$ simultaneously chooses an action $a_{i}\in \mathcal{A}_{i}$ and receives a payoff $u_{i}(a_{i}, \bm{a}_{-i})$ when other agents choose actions $\bm{a}_{-i}$.
\textit{Bayesian games} extend the normal-form games by capturing settings in which agents hold \textit{private information}.
The private information characterizes, e.g., the agent's preference or taste over the outcomes of the game, and determines the payoffs the agent may obtain for every ourcome of the game.
%
%
Unlike normal-form games, each agent $i$ in a Bayesian game does not know the types of all other agents, $\bm{\theta}_{-i}\in \bm{\Theta}_{-i}$.
A common approach to modelling this incomplete information setting is by adopting Harsanyi's idea of introducing a move by the \textit{Nature} \cite{harsanyi1967games}, which handles the agents' uncertainty about others by transforming the incomplete information game into a imperfect information game.
In Harsanyi's model, each agent's private information is known as \textit{type} and is randomly chosen by Nature according to some prior distribution, which is commonly known by all the agents and is referred to as \textit{common prior}.

\begin{definition}[Bayesian Game \cite{osborne2004introduction}] A Bayesian game $\widehat{G}^{B}$ is defined by a tuple $\widehat{G}^{B}\equiv<\mathcal{N}, \bm{\mathcal{A}}, \bm{\Theta}, \bm{d}^{\theta},\bm{u}>$.
%
%
%
$\bm{\Theta}\equiv \times_{i\in \mathcal{N}} \Theta_{i}$, where $\Theta_{i}$ is a type space of agent $i\in \mathcal{N}$; 
$\bm{d}^{\theta} \equiv \{d^{\theta}_{i}\}_{i\in\mathcal{N}}$, where $d^{\theta}_{i}$ is the prior distribution of agent $i$'s type $\theta_{i}\in \Theta_{i}$;
$\bm{u}\equiv \{u_{i}\}_{i\in \mathcal{N}}$, where $u_{i}: \Theta_{i}\times \bm{\mathcal{A}} \mapsto  \mathbb{R}$ is the payoff function of agent $i\in\mathcal{N}$.
\end{definition}

Based on his type, each agent simultaneously takes an action.
%
%
Each agent $i\in\mathcal{N}$ receives a payoff $u_{i}(\theta_{i}, a_{i}, \bm{a}_{-i})$, when his type is $\theta_{i}$, he takes action $a_{i}$ and others $\bm{a}_{-i}$.
We refer to $\bm{\mathcal{O}} \equiv <\bm{\Theta}, \bm{d}^{\theta}>$ as the \textit{global information structure} of the game. 
%
%
%
Let $\bm{\mathcal{O}}_{-i}\equiv <\bm{\Theta}_{-i}, \bm{d}^{\theta}_{-i}>$ denote the information structure of agents other than $i$.
Given a global information structure $\bm{\mathcal{O}}$, we will write the \textit{observation} as $o_{i}\equiv<\theta_{i}| \bm{\mathcal{O}}>$, which contains the information observed by agent $i$.
Here, $\bm{\mathcal{O}}$ is common knowledge and $o_{i}$ is private information of agent $i$ and only partially known (imperfect information) by other agents through $\bm{\mathcal{O}}$.

Markov games generalize normal-form games to dynamic settings as well as Markov decision processing to multi-agent interactions.
A $N$-agent infinite-horizon Markov game is a complete-information game, in which $\bm{\mathcal{O}}=<\mathcal{S}, d^{s}_{0},\mathcal{T}^{s}>$, where 
%
%
$\mathcal{S}$ is a finite set of \text{states}, $d^{s}_{0}$ is the initial distribution of the state, and $\mathcal{T}^{s}:\mathcal{S}\times \bm{\mathcal{A}}\mapsto \Delta(\mathcal{S})$ is the \textit{transition function} of the states; 
all the agents observe the same information, i.e., $o_{i}(s_{t}|\bm{\mathcal{O}}) = o_{j}(s_{t}|\bm{\mathcal{O}})$, for all $i\neq j$.
Each realization of the state, $s_{t}$ is payoff-relevant to all agents and is commonly observed.
The joint actions of agents partially control the dynamics of the states, i.e., the probability distribution of the next state is given by $\mathcal{T}^{s}(\cdot|s_{t}, \bm{a}_{t})$, when the current state is $s_{t}$ and agents take joint actions $\bm{a}_{t}\equiv\{a_{i,t}\}_{i\in\mathcal{N}}\in \bm{\mathcal{A}}$.

\begin{definition}[Markov Game]\label{def:markov_game_original}
A Markov game $\widehat{M}$ is defined by a tuple $\widehat{M}\equiv <\mathcal{N},\bm{\mathcal{A}}, \bm{\mathcal{O}}, \{R_{i}\}_{i\in \mathcal{N}}>$.
$R_{i}: \mathcal{S}\times \bm{A}\mapsto \mathbb{R}$ is a reward function of agent $i\in\mathcal{N}$ that maps state and joint-actions to a reward.
%
\end{definition}

A solution to $\widehat{M}$ is a policy profile $\bm{\pi}:\mathcal{S} \mapsto \Delta(\bm{\mathcal{A}})$, which specifies the joint actions of agents given the state.
In a Markov game, $\bm{\pi}$ can be either independent (i.e., $\bm{\pi}(\bm{a}_{t}| s_{t})=\prod_{i\in \mathcal{N}} \pi_{i}(a_{i,t}|s_{t})$) or correlated (i.e., a joint function).
In this work, we extends Markov games to an incomplete-information setting based on Harsanyi's model.
%
%
The global information structure of the game which is commonly known is $\bm{\mathcal{O}}=<\mathcal{S}, \bm{\Theta}, d^{s}_{0}, \bm{d}^{\theta}_{0}, \mathcal{T}^{s}, \bm{\mathcal{T}}^{\theta}>$,
%
%
where $\bm{\mathcal{T}}^{\theta}\equiv\{\mathcal{T}^{\theta}_{i}\}_{i\in\mathcal{N}}$.
At each period $t\geq 0$, agent $i$ observes $o_{i}=<s_{t}, \theta_{i,t}|  \bm{\mathcal{O}}>$
%
%
, where $s_{t}$ is commonly observed and $\theta_{i,t}$ is the private information of agent $i$ at period $t$.
A special case of imperfect-information Markov game is when agents' private types are static, i.e., $\bm{\mathcal{O}}^{B}=<\mathcal{S}, \bm{\Theta}, d^{s}_{0}, \bm{d}^{\theta}_{0}, \mathcal{T}^{s}>$.
We refer to such Markov game as a \textit{canonical Bayesian Markov game}:


%
\begin{definition}[Canonical Bayesian Markov game]\label{def:canonical_bayesian_markov_game}
A canonical Bayesian Markov game $\widehat{M}^{B}$ is defined by a tuple $\widehat{M}^{B}\equiv<\mathcal{N}, \bm{\mathcal{A}},  \bm{\mathcal{O}}^{B}, \{R_{i}\}_{i\in\mathcal{N}}>$. $R_{i}: \mathcal{S}\times \bm{\mathcal{A}} \times \Theta_{i}\mapsto \mathbb{R}$ is a reward function of agent $i\in\mathcal{N}$ that maps state, joint actions, and his type to a reward.
\end{definition}

A solution to $\widehat{M}^{B}$ is a belief-policy profile $<\bm{\mu}, \bm{\pi}^{B}>$.
Here, $\bm{\mu}\equiv\{\mu_{i}\}_{i\in \mathcal{N}}$ is the \textit{belief system} of the agents and $\mu_{i}\in\Delta(\bm{\Theta}_{-i})$ is each agent $i$'s \textit{belief} about other agents' private information $\bm{\theta}_{-i}$.
The policy profile $\bm{\pi}^{B}: \mathcal{S}\times \bm{\Theta}\mapsto \Delta(\bm{\mathcal{A}})$, such that $\bm{\pi}^{B}(\bm{a}_{t}|s_{t}, \bm{\theta})$ specifies the probability distribution of the joint actions $\bm{a}_{t}$ given the state $s_{t}$ and joint types $\bm{\theta}$.
Similar to the policy profile in $\widehat{M}$, $\bm{\pi}^{B}$ can be either an independent function (i.e., $\bm{\pi}^{B}(\bm{a}_{t}|s_{t}, \bm{\theta}) = \prod_{i\in\mathcal{N}}\pi_{i}^{B}(a_{i,t}|s_{t}, \theta_{i})$) or a correlated function.
We write $\bm{\Pi}$ as the set of policy profiles and $\Pi_{i}$ as a set of policies of agent $i$, for $i\in\mathcal{N}$.

The next step is to define the optimality criteria of $\widehat{M}^{B}$.
Agents' decision makings (i.e., determining $\bm{\pi}^{B}$) are guided by each agent's \textit{expected payoff} (discounted by $0<\gamma\leq 1$).
Specifically, given the global information structure $\bm{\mathcal{O}}^{B}$ and the joint policy, we define agent $i$'s discounted expected payoff as:
\begin{equation}\label{eq:agent_expected_payoff_preliminary}
    \begin{aligned}
    J_{i}(\bm{\pi}^{B}; \mu_{i}|\bm{\mathcal{O}}^{B})\equiv \mathbb{E}^{\mu_{i}}_{\bm{\pi}^{B}}\Big[\sum_{t=0}^{\infty}\gamma^{t} R_{i}(s_{t}, \bm{a}_{t}|\theta_{i})\Big|\bm{\mathcal{O}}^{B}\Big],
    \end{aligned}
\end{equation}
where $\mathbb{E}^{\mu_{i}}_{\bm{\pi}^{B}}\Big[\cdot \Big| \bm{\mathcal{O}}^{B}\Big]$ denotes the expectation with respect to the unique probability law induced by $\bm{\mathcal{O}}^{B}$ and $<\bm{\pi}^{B}, \mu_{i}>$.
We define the notion \textit{Bayesian Markov Nash equilibrium} as a solution concept of $\widehat{M}^{B}$ that extends the concept of \textit{Bayesian Nash equilibrium} \cite{kajii1997robustness} to our imperfect-information Markov setting:

\begin{definition}[BME]\label{def:BMNE_preliminary}
A profile $<\bm{\pi}^{BME}, \bm{\mu}>$ constitutes a Bayesian Markov Nash Equilibrium (BME) if the followings hold, for all $i\in\mathcal{N}$:
\begin{itemize}
    \item[(i)] Optimality: 
    \begin{equation}\label{eq:BMNE_optimality}
        J_{i}(\bm{\pi}^{BME};\mu_{i}|\bm{\mathcal{O}}^{B}) =\sup_{\pi^{B}_{i}\in \Pi_{i}} J_{i}(\bm{\pi}^{BME}_{-i}, \pi^{B}_{i}; \mu_{i}|\bm{\mathcal{O}}^{B}),
    \end{equation}
    where $\bm{\pi}^{BME}_{-i}=\prod_{j\in\mathcal{N}\backslash\{i\}}\pi^{BME}_{j}$;
    \item[(ii)] Consistency: 
    \begin{equation}\label{eq:BMNE_consistency}
    \begin{aligned}
    &\mu_{i}(\bm{\theta}_{-i}|s_{t}, \bm{a}_{-i,t})=\frac{\bm{\pi}^{BME}_{-i}(\bm{a}_{-i,t}| s_{t}, \bm{\theta}_{-i})\bm{d}^{\theta}_{-i,0}(\bm{\theta}_{-i})}{\sum_{s'_{t}}, \bm{\theta}'_{-i,t} \bm{\pi}^{BME}_{-i}(\bm{a}_{-i,t}| s'_{t}, \bm{\theta}'_{-i})\bm{d}^{\theta}_{-i,0}(\bm{\theta}'_{-i,t} )  }.
    \end{aligned}
    \end{equation}
\end{itemize}
\end{definition}
In a BME, the optimality (\ref{eq:BMNE_optimality}) says that each agent $i$'s any deviation from equilibrium $\pi^{BME}$ is not profitable (given all other agents playing equilibrium policy $\bm{\pi}^{BME}_{-i}$).
The consistency (\ref{eq:BMNE_consistency}) requires that in a BME each agent $i$'s belief about other agents' type $\bm{\theta}_{-i}$ to be consistent with the policies played by other agents.

Another solution concept for $\widehat{M}^{B}$ is \textit{Bayesian Markov correlated equilibrium}, which generalizes the BME, such that the equilibrium policy profile $\bm{\pi}$ is a correlated function.
Given $(s_{t}, \bm{a}_{t}, \theta_{i})$, define agent $i$'s interim expected payoff (discounted by $0<\gamma\leq 1$):
\begin{equation}\label{eq:interim_expected_payoff_BMCE}
    \begin{aligned}
    J_{i,t}(s_{t}, \bm{a}_{t}, \theta_{i}; \bm{\pi}|\bm{\mathcal{O}}^{B})\equiv& \mathbb{E}^{\mu_{i}}_{\bm{\pi}}\Big[\sum_{\tau\geq t} \gamma^{\tau}R_{i}(s_{\tau}, \bm{a}_{\tau}|\theta_{i})\Big| \bm{\mathcal{O}}^{B}\Big].
    \end{aligned}
\end{equation}
\begin{definition}[BMCE]\label{def:BMCE}
A profile $<\bm{\pi}^{BMCE}, \bm{\mu}>$ constitutes a Bayesian Markov correlated equilibrium (BMCE) if the followings hold, for all $i\in\mathcal{N}$:
\begin{itemize}
    \item[(i)] Optimality: for any  $s_{t}\in \mathcal{S}$, $a'_{i,t}\in \mathcal{A}_{i}$, $\theta_{i}\in\Theta_{i}$,
    \begin{equation}
        \begin{aligned}
        &\mathbb{E}^{\mu_{i}}_{\bm{a}_{-i,t}\sim\bm{\pi}^{BMCE}_{-i}}\Big[ J_{i,t}(s_{t}, a_{i,t}, \bm{a}_{-i,t},  \theta_{i};\bm{\pi}^{BMCE}|\bm{\mathcal{O}})\Big]
        \geq  \mathbb{E}^{\mu_{i}}_{\bm{a}_{-i,t}\sim\bm{\pi}^{BMCE}_{-i}}\Big[ J_{i,t}(s_{t}, a'_{i,t}, \bm{a}_{-i,t}, \theta_{i};\bm{\pi}^{BMCE}|\bm{\mathcal{O}})\Big],
        \end{aligned}
    \end{equation}
    where $\sum_{\bm{a}'_{-i,t}, \bm{\theta}'_{-i}}\bm{\pi}^{BMCE}(a_{i,t}, \bm{a}'_{-i,t}|s_{t}, \theta_{i},\bm{\theta}'_{-i})>0$.
    \item[(ii)] Consistency:
    \begin{equation}\label{eq:BMCE_consistency}
        \begin{aligned}
          &\mu_{i}(\bm{\theta}_{-i}|s_{t}, \bm{a}_{-i,t}) =\frac{\bm{\pi}^{BMCE}_{-i}(\bm{a}_{-i,t}| s_{t}, \bm{\theta}_{-i})\bm{d}^{\theta}_{-i,0}(\bm{\theta}_{-i})}{\sum_{s'_{t}}, \bm{\theta}'_{-i,t} \bm{\pi}^{BMCE}_{-i}(\bm{a}_{-i,t}| s'_{t}, \bm{\theta}'_{-i})\bm{d}^{\theta}_{-i,0}(\bm{\theta}'_{-i,t} )  },
        \end{aligned}
    \end{equation}
    
    where $\bm{\pi}^{BMCE}_{-i}(\bm{a}_{-i,t}| s_{t}, \bm{\theta}_{-i})$  $=\sum_{a'_{i,t}, \theta'_{i}} \bm{\pi}^{BMCE}($ $a'_{i,t}, \bm{a}_{-i,t}|s_{t}, \theta'_{i},\bm{\theta}_{-i})$.
\end{itemize}
\end{definition}

In BMCEs, agents can coordinate their actions to achieve higher expected payoffs. 
Conceptually, we can imagine that there is a coordinator that uses $\bm{\pi}^{BMCE}$ to provide an action recommendation $a_{i,t}$ specified by $\pi_{i}(a_{i,t}|s_{t}, \theta_{i,t})=\sum_{\bm{a}'_{-i,t}, \bm{\theta}'_{-i}}$ $\bm{\pi}^{BMCE}(a_{i,t}, \bm{a}'_{-i,t}|s_{t},$  $\theta_{i},\bm{\theta}'_{-i})$ to each agent $i$ with belief $\mu_{i}$ about $\bm{\theta}_{-i}$, who knows the distribution of other agents' actions through $\bm{\pi}^{BMCE}_{-i}(\bm{a}_{-i,t}| s_{t}, \bm{\theta}_{-i})$.
To be an equilibrium, $\bm{\pi}^{BMCE}$ is required to incentivize each agent $i$ to take the recommended action, instead of deviating to another action.
However, similar to standard correlated equilibrium, this coordinator is not required to achieve a BMCE as long as there is a public communication mechanism, e.g., publicly observed information \cite{ziebart2011maximum,dodis2000cryptographic}.


\section{The Information Design Problem}\label{sec:the_information_design_problem}

Consider a discrete-time $n$-agent information-horizon game $M$ that extends a canonical Bayesian Markov game $\widehat{M}^{B}$ by expanding the information structure with additional payoff-relevant information, referred to as \textit{signals}.
We refer to $M$ as the \textit{augmented Bayesian Markov game} (A-BMG, augmented game).
We consider an environment in which there are $m$ sources of signals, denoted as $\mathcal{K}\equiv[m]$. Each source $l\in \mathcal{K}$ of signals sends one signal $\omega^{l}_{i,t}$ to each agent $i$ at each period $t$ of the game.
%
%
We refer to the commonly observed state as \textit{global state} (state) $g_{t}$.
Hence, besides a state $g_{t}$ and private type $\theta_{i}$, each agent $i$ privately observes a group of signals, denoted by $W_{i,t}\equiv\{\omega^{l}_{i,t}\}_{l\in \mathcal{K}}$.
However, each agent $i$ selects only one signal $\omega_{i,t}$ from the group $W_{i,t}$.
The game $M$ is defined as a tuple:
\begin{equation}\label{tuple:Bayesian_Markov_game}
    M\equiv<\mathcal{N}, \mathcal{K}, \mathcal{A},   \bm{\mathcal{O}}, \{R_{i}\}_{i\in \mathcal{N}}>.
\end{equation}
Here, $\mathcal{A}$ is a finite set of \textit{actions} each agent can choose from; let $\bm{\mathcal{A}}\equiv \mathcal{A}^{n}$ denote the set of joint actions and $\bm{A}_{-i}\equiv \mathcal{A}^{n-1}$ denote the set of joint actions of agents other than $i$.
$\bm{\mathcal{O}}\equiv<\mathcal{G}, \Omega^{m}, \bm{\Theta}, \mathcal{T}_{g}, \mathcal{P}, d_{g}, d_{\theta}>$ is the global information structure, where where $\mathcal{G}$ is a finite set of \textit{global states} (states); $\Omega$ is a finite set of \textit{signals} each source from $\mathcal{K}$ can send;
$\bm{\Theta}\equiv \Theta^{n}$, where $\Theta$ is a finite set of \textit{types} and each agent $i$'s type $\theta_{i}$ is privately observed by agent $i$ (we assume that agents have the same set of types); $\mathcal{T}_{g}: \mathcal{G}\times \bm{\mathcal{A}} \mapsto \Delta(\mathcal{G})$
is a \textit{transition function} of the state, such that $\mathcal{T}_{g}(g_{t+1}|g_{t}, \bm{a}_{t})$ specifies the probability of the next state $g_{t+1}$ when the current global state is $g_{t}$ and current joint actions are $\bm{a}_{t}$;
%
%
$\mathcal{P}$ is the \textit{probability measure} of the signals $\bm{W}_{t}\equiv\{W_{i,t}\}_{i\in\mathcal{N}}$ received by the agents as additional information;
$d_{g}$ is the \textit{initial distribution} of the state;
$d_{\theta}$ is the \textit{prior distribution} of each agent's type.
After receiving $W_{i,t}$, agent $i$ selects one signal $\omega_{i,t}$ from $W_{i,t}$.
$R_{i}: \mathcal{G} \times \mathcal{S} \times \bm{\mathcal{A}} \times \Omega \times \Theta \mapsto \mathbb{R}$ is the \textit{reward function} that maps  the joint actions $\bm{a}_{t}\in \bm{\mathcal{A}}$, global state $g_{t}$, agent $i$'s selected signal $\omega_{i,t}$, and his type $\theta_{i}$ into a scalar reward $R_{i}(\bm{a}_{t}, g_{t}, \omega_{i,t}|\theta_{i})$.

Each agent $i$ is \textit{rational} in the sense that it is self-interested and makes his decisions according to his observation $o= (g_{t}, W_{i,t}, \theta_{t}|\bm{\mathcal{O}})$ to maximize his expected payoffs. 
Here, each agent $i$ privately observes his type $\theta_{i}$ and signals $W_{i,t}$. Hence, only the global state $g_{t}$ is the common information among the observations of the agents.
Given the observation $o_{i} = (g_{t}, W_{i,t}, \theta_{i}|\bm{\mathcal{O}})$, the decision making of each agent $i$ consists of two processes: \textit{(i)} selecting one signal $\omega_{i,t}$ from $W_{i,t}$ and \textit{(ii)} choosing an action $a_{i,t}$ from $\mathcal{A}$. 
The solution to the game $M$ is a profile $<\bm{\beta}, \bm{\pi}, \bm{\mu}>$.
Here, $\bm{\beta}: \mathcal{G} \times \Omega^{m\times n} \times \bm{\Theta} \mapsto \bm{\Omega}$ is a \textit{selection strategy profile}, such that $\bm{\beta}(g_{t}, \bm{W}_{t}, \bm{\theta})$ specifies their choices of signals $\bm{\omega}_{t}\equiv \{\omega_{i,t}\}_{i\in N}$ for each observation profile $\bm{o} = (g_{t}, \bm{W}_{t}, \bm{\theta}_{t}|\bm{\mathcal{O}})$, and 
$\bm{\pi}: \mathcal{G}\times \Omega^{n} \times \bm{\Theta}\mapsto \Delta(\bm{\mathcal{A}})$ is a \textit{policy profile}, such that $\bm{\pi}(\bm{a}_{t}|g_{t}, \bm{\omega}, \bm{\theta})$ specifies the distribution of the next joint actions, for each state $g_{t}$, choice of joint signals $\bm{\omega}_{t}$, and joint types $\bm{\theta}$.
The profiles $\bm{\beta}$ and $\bm{\pi}$ can be either \textit{correlated} (i.e, a joint function) or \textit{independent} (i.e.,  $\omega_{i,t}=\beta_{i}(g_{t}, W_{i,t}, \theta_{i})$, for all $i\in \mathcal{N}$, and $\bm{\pi} = \prod_{i\in N} \pi_{i}$).
The solution of the augmented game $M$ also requires a \textit{belief} system $\bm{\mu}=\{\mu_{i}\}_{i\in \mathcal{N}}$, where $\mu_{i}: \mathcal{G} \times \Omega\times \Theta \mapsto \Delta(\Omega^{(n-1)}\times \Theta^{n-1})$ which describes each agent $i$'s belief about unobserved signals $\bm{W}_{-i}$ of other agents and and the unobserved types of other agents $\bm{\theta}_{-i}$.
%
%

Given any observation $o_{i}=(g_{t}, W_{i,t}, \theta_{i}|\bm{\mathcal{O}})$, each agent $i$'s selection of the signal and the choice of action are fundamentally different.
Specifically, agent $i$ first uses $\beta_{i}$ to select signal $\omega_{i,t}=\beta_{i}(g_{t}, W_{i,t}, \theta_{i})$ and then chooses an action $a_{i,t}$ according to $\pi_{i}(a_{i,t}|g_{t}, \omega_{i,t}, \theta_{i})$ (suppose we consider a Nash equilibrium here) based on the realized selection $\omega_{i,t}$. The transition of the global state is controlled by the current $g_{t}$ and the realized actions $\bm{a}_{t}$, i.e., $\mathcal{T}_{g}(g_{t+1}|g_{t}, \bm{a}_{t})$ and, however, is independent of the selected signal $\omega_{i,t}$, for all $i\in\mathcal{N}$, given $g_{t}, \bm{a}_{t}$.

In this work, we are interested in that there is one rational information designer referred to as \textit{principal} (she, indexed as $k$) that controls one of $m$ sources of signals.
The principal privately sends a signal $\omega^{k}_{i,t}$ to each agent $i$ such that $\bm{\omega}^{k}_{t}$ is distributed according to some probability measure $\mathcal{P}^{k}\in \Delta(\Omega^{n})$.
We assume that $\bm{W}^{-k}_{t}\equiv \bm{W}_{t}\backslash\{\bm{\omega}^{k}_{t}\}$ is distributed according to some fixed
$\mathcal{P}^{-k}$.
%
%
%
%
%
%
%
We consider that the principal is \textit{rational} in the sense that she strategically chooses $\mathcal{P}^{k}$ that governs the realizations of the her signals $\bm{\omega}^{k}_{t}$ at each period $t$, thereby influence $\mathcal{P}$ of $\bm{W}_{t}=\{\bm{\omega}^{k}_{t}, \bm{W}^{-k}_{t}\}$, such that the equilibrium behaviors of the agents coincide with the principal's desired equilibrium.
This process is \textit{information design:}

\begin{definition}[\textbf{Information Design Problem}]
An information design problem is defined as a tuple $\mathcal{I}\equiv<M, \bm{\pi}, \{\mathcal{P}^{k}_{i}\}_{i\in N}, \Omega, \bm{\kappa}>$. 
Here, $M$ is a Bayesian Markov game model defined by (\ref{tuple:Bayesian_Markov_game}); $\bm{\pi}$ is the agents' policy profile; 
$<\mathcal{P}^{k}_{i}, \Omega>$ is the information structure, where $\mathcal{P}^{k}_{i}$ defines a distribution of the signal $\omega^{k}_{i,t}$ sent by the principal at each $t$;
$\bm{\kappa}: \mathcal{G} \times \bm{\Theta}\mapsto \Delta(\bm{A})$ is the principal's target equilibrium probability distribution of agents' joint action conditioning only on the state and agents' type.
%
%
\end{definition}

A solution to $\mathcal{I}$ is a \textit{signaling rule profile}
$\bm{\alpha}: \mathcal{G}\times \bm{\Theta}\mapsto \Delta(\bm{\Omega})$ that defines $\bm{\mathcal{P}}^{k}\equiv \{\mathcal{P}^{k}_{i}\}_{i\in N}$ of the joint signal $\bm{\omega}^{k}_{t}$, i.e., $\bm{\alpha}(\bm{\omega}^{k}_{t}|g_{t}, \bm{\theta})$ specifies the probability distribution of $\bm{\omega}^{k}_{t}$ sent by the principal to the agents at period $t$, when the state is $g_{t}$ and the joint types are $\bm{\theta}$.
We will write the augmented game $M$ with the principal using $\bm{\alpha}$ as $M[\bm{\alpha}]$ and refer to it as \textit{$\bm{\alpha}$-augmented game}.
If the information design is viewed as an extensive form game between the principal and the agents, the timing is as follows\footnote{This extensive form game is different from the Markov game $M[\bm{\alpha}]$ and is described for the purpose of timing the decision making processes of the principal and the agents. The principal is not a player of $M[\bm{\alpha}]$.}:
\begin{itemize}
    \item[(i)] the principal chooses a signaling rule profile $\bm{\alpha}$ for the agents, each of whom has a privately realized type $\theta_{i}$;
    \item[(ii)] a state $g_{t}$ is realized at the beginning of each period $t$ and is observed by the principal and all the agents;
    \item[(iii)] the principal privately sends $\omega^{k}_{i,t}$ to each agent $i$ and each agent $i$ receives $W_{i,t}=\{\omega^{k}_{i,t}, W^{-k}_{i,t}\}$;
    %
    \item[(iv)] the agents chooses their signals $\bm{\omega}_{t}$ (i.e., the selections) from $\bm{W}_{t}$ according to $\bm{\beta}$;
    \item[(v)] the agents chooses their actions according to $\bm{\pi}$ based on their private types, state, and signal;
    \item[(vi)] immediate rewards are realized and the state $g_{t}$ is transitioned to $g_{t+1}$ according to $\mathcal{T}$.
\end{itemize}

Given an $\bm{\alpha}$, we re-write the global information structure of $M[\bm{\alpha}]$ as $\bm{\mathcal{O}}^{\bm{\alpha}}=<\mathcal{G}, \Omega^{m}, \bm{\Theta}, \mathcal{T}_{g}, \mathcal{P}^{-k}, \bm{\alpha}, d_{g}, d_{\theta}>$ and an observation as $o= (g_{t}, \omega^{k}_{i,t}, W^{-k}_{i,t}, \theta_{t}|\bm{\mathcal{O}})$.
Here, we assume that $\bm{\mathcal{O}}^{\bm{\alpha}}$ is common knowledge (i.e., common prior) and only the realizations of $\theta_{i}$, $\omega^{k}_{i,t}$, and $W^{-k}_{i,t}$, for $t\geq 0 $ are unobserved by agents other than $i$. 
Hence, each agent $i$ is a Bayesian decision maker.
%
%

\subsection{Implementability}

According to Ionescu Tulcea theorem (see, e.g., Hern$\acute{\text{a}}$ndez-Lerma Lasserre \cite{hernandez2012discrete}), initial distribution $d_{g}$ on $g_{0}$, transition function $\mathcal{T}_{g}$, signaling rule profile $\bm{\alpha}$ and distribution $\mathcal{P}^{-k}$, selection rule profile $\bm{\beta}$, policy profile $\bm{\pi}$ define a unique probability measure $P^{\bm{\pi}, \bm{\beta}, \bm{\alpha}}$ on $(\mathcal{G} \times \Omega^{m\times n} \times \bm{\mathcal{A}})^{\infty}$. 
Given the belief $\mu_{i}$, the expectation with respect to $P^{\bm{\pi}, \bm{\beta}, \bm{\alpha}}$ denoted by $\mathbb{E}^{\bm{\beta},\bm{\alpha};\mu_{i}}_{\bm{\pi}}\big[ \cdot \big]$ or $\mathbb{E}^{\bm{\beta},\bm{\alpha};\mu_{i}}_{\bm{\pi}}\big[ \cdot \big| \cdot\big]$. 
Here, since the global information structure $\bm{\mathcal{O}}^{\bm{\alpha}}$ is fixed except the signaling rule $\bm{\alpha}$, we show the $\bm{\alpha}$ instead of $\bm{\mathcal{O}}^{\bm{\alpha}}$ in the notations of expectation.
Each agent $i$'s decision making is governed by its (discounted, $0<\gamma\leq 1$) \textit{cumulative expected reward} (expected reward):
%
\begin{equation}\label{eq:CER}
    \begin{split}
    &\mathtt{ExpR}_{i}^{\bm{\pi},\bm{\beta}, \bm{\alpha}; \mu_{i}}(\bm{a}_{t}, g_{t}, \omega_{i,t} ; \{\omega^{k}_{i,t}, W^{-k}_{i,t}\}| \theta_{i})
    \equiv \sum_{\bm{\theta}_{-i}} \mathbb{E}^{\bm{\beta}, \bm{\alpha};\mu_{i}}_{\bm{\pi}}
    \Big[ \sum_{\tau \geq t} \gamma^{\tau}R_{i}(\tilde{\bm{a}}_{\tau}, \tilde{g}_{\tau},  \tilde{\omega}_{i,\tau}|\theta_{i}) \Big|g_{t}, W_{i,t}, \bm{\theta}_{-i}\Big]\bm{d}_{\theta}(\bm{\theta}_{-i}),
    \end{split}
\end{equation}
where
%
%
$\omega_{i,t} = \beta_{i}(g_{t}, W_{i,t},\theta_{i})$ and the period-$\tau$ selected signal $\tilde{\omega}_{i,\tau}$ is a random variable whose distribution is determined by $\alpha_{i}$, $\mathcal{P}^{-k}$, and $\beta_{i}$, and $\bm{d}_{\theta}(\bm{\theta}_{-i}) = \prod_{j\neq i } d_{\theta}(\theta_{j})$.
Agents choose $\bm{\beta}$ and $\bm{\pi}$ by maximizing the expected reward (\ref{eq:CER}).
For notational simplicity, we will remove $W^{-k}_{i,t}$ from $W_{i,t}$ and only show $\omega^{k}_{i,t}$, unless otherwise stated, e.g., $\mathtt{ExpR}_{i}^{\bm{\pi},\bm{\beta}, \bm{\alpha}; \mu_{i}}(\bm{a}_{t}, g_{t}, \omega_{i,t} ; \omega^{k}_{i,t}| \theta_{i}) = $ $\mathtt{ExpR}_{i}^{\bm{\pi},\bm{\beta}, \bm{\alpha}; \mu_{i}}(\bm{a}_{t}, g_{t}, \omega_{i,t} ; \{\omega^{k}_{i,t}, W^{-k}_{i,t}\}| \theta_{i})$.
%

%
As in a standard Markov game, each agent $i$'s decision of choosing an action $a_{t}$ takes into account other agents' decisions of choosing $\bm{a}_{-i,t}$ because its immediate reward of taking $a_{i,t}$ directly depends on $\bm{a}_{-i,t}$.
In $M[\bm{\alpha}]$, agent $i$'s choices of $\beta_{i}$ and $\pi_{i}$ are coupled because $\omega_{i,t}$ specified by $\beta_{i}$ has a direct causal effect on $a_{i,t}$ through $\pi_{i}$.
Thus, other agents' immediate reward indirectly depends on each individual agent's selected signal through his action.
Hence, agents' strategic interactions in $M[\bm{\alpha}]$ consist of selecting signals by $\bm{\beta}$ and taking actions by $\bm{\pi}$.
Since $\mathcal{P}^{-k}$ is fixed, the principal's choice of $\bm{\alpha}$ controls the dynamics of $\bm{W}_{t}$.
Therefore, it is possible for the principal to influence the equilibrium behaviors of agents in the game $M[\bm{\alpha}]$ through proper designs of $\bm{\alpha}$.

The principal's information design problem $\mathcal{I}$ is a mechanism design problem that takes an objective-first approach to design information structures of signals sent to agents, toward desired objectives $\bm{\kappa}$, in a strategic setting through the design of $\bm{\alpha}$, where self-interested agents act rationally by choosing $\bm{\beta}$ and $\bm{\pi}$.
The choice of $\bm{\alpha}$ is independent of the realizations of states and agents' types.
The key restriction on the principal's $\bm{\alpha}$ is that the agents are elicited to perform equilibrium behaviors $\bm{\pi}$ that coincides with the principal's desired equilibrium $\bm{\kappa}$.
This is captured by a notion of \textit{implementability}:

\begin{definition}[\textbf{Implementability}]\label{def:implementability}

Given $\bm{\kappa}$, $M$, and $\mathcal{P}^{-k}$, the signaling rule profile $\bm{\alpha}$ is \textbf{implementable} if it elicits a strategy profile $<\bm{\beta}^{*}, \bm{\pi}^{*}>$ that is an equilibrium and $\bm{\pi}^{*}$ is \textbf{admissible} equilibrium $\bm{\pi}^{*}$ of $M[\bm{\alpha}]$, i.e., 
for all $i\in N$, $t\in\mathbb{T}$, $(g_{t}, \theta_{i}, \bm{\theta}_{-i})\in \mathcal{G}\times \bm{\Theta }$,
    \begin{equation}\label{eq:def_admissible}
    \begin{aligned}
       \sum_{ \bm{\omega}^{k}_{t}, \bm{W}^{-k}_{t} } \bm{\pi}^{*}\big(\bm{a}_{t}|g_{t}, \bm{\beta}^{*}(g_{t}, \bm{\omega}^{k}_{t}, \bm{\theta}), \bm{\theta}\big)\bm{\alpha}(\bm{\omega}^{k}_{t}|g_{t}, \bm{\theta})
       \times \mathcal{P}^{-k}(\bm{W}^{-k}_{t}) = \bm{\kappa}(\bm{a}_{t}| g_{t}, \bm{\theta}).
       \end{aligned}
    %
    \end{equation}
%
%
In this case, we say that agents' equilibrium $<\bm{\beta}^{*}, \bm{\pi}^{*}>$ \textbf{implements} $\bm{\alpha}$.
\end{definition}


%
Given any $\mathcal{P}^{-k}$, the distribution of $\bm{a}$ conditioning on any state $g$ is jointly determined by the agents' $\bm{\beta}$ and the principal's $\bm{\alpha}$.
Hence, given $\mathcal{P}^{-k}$, the signal $\bm{\omega}^{k}_{t}$ sent by the principal by using $\bm{\alpha}$ ultimately influences each agent' expected reward. However, this information is transmitted \textit{indirectly} through the agents' selection rules, i.e., $\bm{\omega}_{t} = \bm{\beta}(g_{t}, \bm{\omega}^{k}_{t}|\bm{\theta})$, where $\bm{\omega}_{t}$ is not necessarily equal to $\bm{\omega}^{-k}_{t}$.
Therefore, the information design problem $\mathcal{I}$ aiming to find implementable $\bm{\alpha}$ leads to a \textit{indirect information design} (IID). 
%
We will call the game $M^{-D}[\bm{\alpha}]$ as \textit{indirect augmented game}.


\subsection{Direct Information Design}\label{sec:direct_IDP}

As a designer, the principal takes into consideration each agents decision makings at all aspects of the game.
Hence, in any indirect augmented game $M^{-D}[\bm{\alpha}]$, the principal's design of $\bm{\alpha}$ must predict the possible equilibrium selection rule profile $\bm{\beta}$ and the corresponding equilibrium $\bm{\pi}$ that might be indirectly induced by $\bm{\alpha}$. 
In contrast to $M^{-D}[\bm{\alpha}]$, the principal may elect to a \textit{direct information design}.
%
%
In direct information design, the principal manipulates the agents' equilibrium by directly touching their policies, when each agent selects the signal sent by the principal at each state.
\begin{definition}[\textbf{Direct Information Design}]\label{def:direct_information_design}  
A direct information design induces a \textit{direct augmented game} $M^{D}[\bm{\alpha}]$, in which agents select $\bm{\omega}^{k}_{t}$ sent by $\bm{\alpha}^{D}$ at every period, i.e., $\bm{\beta}(g_{t}, \{\bm{\omega}^{k}_{t}, \bm{W}^{-k}_{t}\},\bm{\theta}) = \bm{\omega}^{k}_{t}$, for all $g_{t}\in \mathcal{G}$, $\{\bm{\omega}^{k}_{t}, \bm{W}^{-k}_{t}\}\in \Omega^{m\times n}$, $\bm{\theta}\in \bm{\Theta}$, $t\in \mathbb{T}$.
\end{definition}

In a $M^{D}[\bm{\alpha}]$, the principal wants $\bm{\omega}^{k}_{t}$ to directly enter the immediate rewards of the agents at each period $t$. 
The key constraint on the $\mathcal{I}$ of $M[\bm{\alpha}]^{D}$ is a notion of \textit{obedient implementability} that requires the design of $\bm{\alpha}$ to be such that \textit{(i)} agents would want to select the signal $\bm{\omega}^{k}_{t}$ sent by the principal than choose any other signals from $\bm{W}^{-k}_{t}$, for each $t\in\mathbb{T}$, and \textit{(ii)} agents take actions specified by the admissible policy other than other available actions.

\begin{definition}[OIL]\label{def:O_Implementability}
Given $\bm{\kappa}$, the signaling rule $\bm{\alpha}^{OIL}$ is \textbf{obedient}\textbf{-implementable} (OIL, obedient-implementability) if it induces $<\bm{\beta}^{O}, \bm{\pi}^{O}>$, such that
\begin{itemize}
    \item[(i)]  $\bm{\beta}^{O}$ is \textbf{obedient}, i.e., for all $i\in N$, $t\in \mathbb{T}$, $(g_{t}, \theta_{i}, \{\omega^{k}_{i,t}, W^{-k}_{i,t}\})\in \mathcal{G} \times \Theta \times \Omega^{m}$,
    \begin{equation}
        \begin{split}
            \bm{\beta}^{O}(g_{t}, \{\bm{\omega}^{k}_{t}, \bm{W}^{-k}_{i,t}\},\bm{\theta}) = \bm{\omega}^{k}_{t};
        \end{split}
    \end{equation}
    \item[(i)] $\bm{\pi}^{O}$ is \textbf{admissible}, i.e., for all $i\in N$, $t\in\mathbb{T}$, $(g, \theta_{i}, \bm{\theta}_{-i})\in \mathcal{G}\times \bm{\Theta }$, $\bm{\pi}^{O}$ is an equilibrium and
    \begin{equation}
        \sum_{\bm{\omega}^{k}} \bm{\pi}^{O}(\bm{a}|g, \bm{\omega}^{k}, \bm{\theta}) \bm{\alpha}^{OIL}(\bm{\omega}^{k}|g, \bm{\theta}) = \bm{\kappa}(\bm{a}| g, \bm{\theta}).
    \end{equation}
\end{itemize}
\end{definition}

Next, we introduce the \textit{obedient perfect Bayesian Markov Nash equilibrium} (O-PBME) as the equilibrium solution concept for the direct augmented game $M^{D}[\bm{\alpha}]$.
%
%
Given $\bm{\beta}$ with $\omega_{i,t} = \beta_{i}(g_{t}, \omega^{k}_{i,t}|\theta_{i})$ and $\bm{\omega}_{-i,t} = \beta_{i}(g_{t}, \bm{\omega}^{k}_{-i,t}|\bm{\theta}_{-i})$ and each agent $i$'s belief $\mu_{i}$, define 
\begin{equation}\label{eq:policy_others_i}
    \begin{aligned}
    \bm{\pi}_{[-i]} (\bm{a}_{-i,t}|g_{t}, \omega_{i}, \theta_{i};\mu_{i}, \bm{\beta})
\equiv \sum_{\bm{\omega}_{i,t}, \bm{\theta}_{-i}} \prod_{j\neq i} \pi_{j}(a_{j,t}| g_{t}, \beta_{j}(g_{t}, \omega^{k}_{j,t},\theta_{j}), \theta_{j})\mu_{i}(\bm{\omega}^{k}_{-i,t}, \bm{\theta}_{-i}|g_{t}, \omega^{k}_{t})\bm{d}_{\theta}(\bm{\theta}_{-i}),
\end{aligned}
\end{equation}
%
%
such that $\bm{\pi}_{[-i]}$ is the joint policy of all other agents as perceived by agent $i$ given $\mu_{i}$ and $\bm{\beta}$, which is independent of $a_{i,t}$.
The optimality criterion of each agent $i$ is captured by his expected payoff:
\begin{equation}\label{eq:MBCE_EP}
    \begin{aligned}
    \texttt{ExE}^{\bm{\pi}, \bm{\beta}, \bm{\alpha};\mu_{i} }_{i} &(a_{i,t}; g_{t}, \omega_{i,t}; \omega^{k}_{i,t}|\theta_{i})\equiv \mathbb{E}_{\tilde{\bm{a}}_{-i,t}\sim \bm{\pi}_{[-i]}(\tilde{\bm{a}}_{-i,t}|g_{t}, \omega_{i}, \theta_{i}; \mu_{i}, \bm{\beta})} \Big[\mathtt{ExpR}_{i}^{\bm{\pi},\bm{\beta}, \bm{\alpha}; \mu_{i}}(a_{i,t}, \tilde{\bm{a}}_{t}, g_{t}, \omega_{i,t}; \omega^{k}_{i,t}|\theta_{i}) \Big].
    \end{aligned}
\end{equation}
%
%
%
%
The belief system $\bm{\mu}$ is \textit{consistent} if it is updated according to the Bayes' rule:
\begin{equation}\label{eq:belief}
    \begin{aligned}
    &\mu_{i}( \bm{\omega}_{-i,t}, \bm{\theta}_{-i} |g_{t},W_{i,t},\theta_{i};\bm{\beta}_{-i})
    = \frac{\bm{\alpha}(\omega^{k}_{i,t}, \bm{\omega}^{k}_{-i, t}| g_{t}, \{\theta_{i}, \bm{\theta}_{-i}\}) \bm{\mathcal{P}}^{-k}( \bm{W}^{-k}_{-i, t})\bm{d}_{\theta}(\bm{\theta}_{-i})}{\sum\limits_{\bm{\hat{W}}_{-i,t}, \hat{\bm{\theta}}_{-i} } \bm{\alpha}(\omega^{k}_{i,t}, \bm{\hat{\omega}}^{k}_{-i, t}| g_{t}, \{\theta_{i}, \bm{\theta}_{-i}\}) \bm{\mathcal{P}}^{-k}( \bm{\hat{W}}^{-k}_{-i, t}) \bm{d}_{\theta}(\hat{\theta}_{-i})}.
    \end{aligned}
\end{equation}

%
%
When the denominator of (\ref{eq:belief}) is zero, then agent $i$ sets any probabilistic belief about $<\bm{\omega}_{-i,t},\bm{\theta}_{-i}>$.
%
We formally define the O-PBME as follows.

\begin{definition}[O-PBME]\label{def:PBE_markov}
A profile $<\bm{\beta}^{*}, \bm{\pi}^{*}>$ with $\bm{\mu}$ constitutes a PBME if the belief $\bm{\mu}$ is updated according to (\ref{eq:belief}) and the the policy profile is independent, i.e., $\bm{\pi}^{*}(\bm{a}_{t}|g_{t}, \bm{\omega}_{t}, \bm{\theta})$  $=\prod_{i\in\mathcal{N}}$ $\pi^{*}_{i}(a_{i,t}| g_{t}, \omega_{i,t}, \theta_{i})$, such that, for any $g_{t}\in\mathcal{G}$, $\theta_{i}\in\Theta$, $\omega^{k}_{i,t}\in \Omega$ with $\alpha_{i}(\omega^{k}_{i,t}|g, \theta_{i})>0$, $\omega_{i,t} = \beta_{i}(g_{t}, \omega^{k}_{i,t}, \theta_{i})$, $\omega'_{i,t}\in \Omega$, $a_{i,t}\in \mathcal{A}$ with $\pi^{*}_{i}(a_{i,t}| g_{t}, \omega_{i,t}, \theta_{i})>0$, $a'_{i,t}\in \mathcal{A}$, $i\in\mathcal{N}$,
\begin{equation}\label{eq:def_O_PBME}
    \begin{aligned}
    \mathtt{ExE}^{\bm{\pi}^{*}, \bm{\beta}^{*}, \bm{\alpha};\mu_{i} }_{i} &(a_{i,t}; g_{t}, \omega_{i,t};\omega^{k}_{i,t}|\theta_{i})
    \geq \mathtt{ExE}^{\bm{\pi}^{*}, \bm{\beta}, \bm{\alpha};\mu_{i} }_{i} (a'_{i,t}; g_{t}, \omega'_{i,t}; \omega^{k}_{i,t}|\theta_{i}).
    \end{aligned}
\end{equation}
The PBME profile is an O-PBME if the selection rule profile $\bm{\beta}^{*}$ is obedient.
%
%
We denote O-PBME equilibrium profile as $<\bm{\beta}^{O}, \bm{\pi}^{\text{O}}>$ with $\bm{\mu}$ and refer to the signaling rule as OIL in O-PBME (OIL-P), denoted as $\bm{\alpha}^{\text{OIL}}$, if it induces an O-PBME in which the equilibrium policy profile $\bm{\pi}^{\text{OP}}$ is admissible, denoted as $\bm{\pi}^{\text{AO}}$. We call $<\bm{\beta}^{O}, \bm{\pi}^{\text{AO}}>$ as an admissible O-PBME.
\end{definition}

A successful information design depends on the principal's having accurate beliefs in regard to the agents' decision processes. This includes all the possible indirect selection behaviors of the agents, i.e., all possible $\bm{\beta}\neq \bm{\beta}^{O}$. 
The point of direct information design is that it allows the principal to ignore analyzing all of agents' indirect selections behaviors and focus on the obedient $\bm{\beta}^{O}$.
This is promoted by the \textit{obedience principle:}

\begin{theorem}[\textbf{Obedience Principle}]\label{theorem:obedience_principle}
Let $<\bm{\beta}^{*}, \bm{\pi}^{*}>$ with $\bm{\mu}$ implement an indirect $\bm{\alpha}^{-D}$ in a PBME that achieves a goal $\bm{\kappa}$.
Then, there exists a direct information design with OIL-implementable $\bm{\alpha}^{OIL}$ in PBME that induces an equilibrium $<\bm{\beta}^{O}, \bm{\pi}^{\text{AO}}>$ of the game $M[\bm{\alpha}^{OIL}]$ that achieves the same goal $\bm{\kappa}$.

\end{theorem}

The obedience principle shows that for any indirect information design that achieves a goal, there exists a direct information design that leads to the same goal.
Hence, it is without loss of generality for the principal to focus on direct information design, in which the agents' obedient selection strategy $\bm{\beta}^{O}$ is straightforward.

\section{Obedient Implementability}\label{sec:obedient_implementability}

In this section, we characterize the OIL-P of the signaling rule $\bm{\alpha}^{OIL}$ that constrains the principal's information design problem.
Since we restrict attention on stationary equilibrium strategies, we omit the time index unless otherwise states.
%

\subsection{Equilibrium Analysis of \texorpdfstring{$M[\bm{\alpha}]$}{TEXT}}

Given a game $M[\bm{\alpha}]$ for some $\bm{\alpha}$, let $\mathcal{T}^{\bm{\pi}, \bm{\beta}, \bm{\alpha} }_{g'g}$ denote the transition probability from state $g$ to state $g'$, given that agents chooses action according to $\bm{\pi}$ and selects signals according to $\bm{\beta}$, with a slight abuse of notation:
$$
\mathcal{T}^{\bm{\pi}, \bm{\beta}, \bm{\alpha}}_{g'g} \equiv \sum_{\bm{a}, \bm{\omega}^{k}, \bm{\omega}}\bm{\pi}(\bm{a}|g, \bm{\omega}, \bm{\theta})\bm{\alpha}(\bm{\omega}^{k}|g, \bm{\theta}) \mathcal{T}_{g}(g'|g, \bm{a}).
$$
Given $\bm{\alpha}$, $\bm{\beta}$, $\bm{\pi}$, and the belief system $\bm{\mu}$, define the \textit{state-signal value} function $V^{\bm{\pi}, \bm{\beta}, \bm{\alpha}; \mu_{i}}$ of agent $i$, representing agent $i$'s expected reward, originating at $(g, \omega^{k}_{i}, \omega_{i})\in \mathcal{G}\times \Omega \times \Omega$ with $\omega_{i} = \beta_{i}(g, \omega^{k}_{i} |\theta_{i})$:
%
%
\begin{equation}\label{eq:state_signal_value_function}
    \begin{aligned}
    &V^{\bm{\pi}, \bm{\beta}, \bm{\alpha}; \mu_{i}}_{i} (g, \omega_{i}; \omega^{k}_{i}|\theta_{i}) \equiv \mathbb{E}^{\bm{\beta};\mu_{i}}\Big[\sum_{t=0}^{\infty}\sum_{g'}
   \gamma^{t}\big(\mathcal{T}^{\bm{\pi}, \bm{\beta}, \bm{\alpha}}_{g'g}\big)^{t}\sum_{\bm{a}}  \bm{\pi}(\bm{a}|g, \bm{\omega}; \bm{\theta})\bm{\alpha}(\bm{\omega}^{k}|g,\bm{\theta})R_{i}(\bm{a}, g, \omega_{i}|\theta_{i}) \Big].
    \end{aligned}
\end{equation}
%
Define the \textit{state value function} $J^{\bm{\pi}, \bm{\beta}, \bm{\alpha}; \mu_{i}}_{i}$ of agent $i$ that describes his expected reward, originating at state $g\in \mathcal{G}$:
%
\begin{equation}\label{eq:state_value_function}
    \begin{aligned}
    &J^{\bm{\pi}, \bm{\beta}, \bm{\alpha}; \mu_{i}}_{i} (g |\theta_{i}) 
    \equiv  \mathbb{E}^{\bm{\beta};\mu_{i}}\Big[\sum_{t=0}^{\infty}\sum_{g'}\ \gamma^{t}\big(\mathcal{T}{}^{\bm{\pi}, \bm{\beta}, \bm{\alpha}}_{g'g}\big)^{t}\sum_{\bm{a}, \bm{\omega}^{k}, \bm{\omega}}  \bm{\pi}(\bm{a}|g, \bm{\omega}; \bm{\theta})\bm{\alpha}(\bm{\omega}^{k}|g,\bm{\theta})R_{i}(\bm{a}, g, \omega_{i}|\theta_{i}) \Big].
    \end{aligned}
\end{equation}

Next, define the \textit{$Q$-value function} (the state-signal-action value function) $Q^{\bm{\pi}, \bm{\beta}, \bm{\alpha}; \mu_{i}}_{i}$ that represents agent $i$'s expected reward if $(\bm{\omega}, \bm{a})\in \Omega^{n} \times \bm{\mathcal{A}}$ are played in $(g, \bm{\omega}^{k})\in \mathcal{G}\times \Omega^{n}$:
%
\begin{equation}\label{eq:agents_Q_function}
    \begin{aligned}
    &Q^{\bm{\pi}, \bm{\beta}, \bm{\alpha}; \mu_{i}}_{i}  (g, \omega_{i}, \bm{a};\omega^{k}_{i}|\theta_{i})
    \equiv \mathbb{E}^{\bm{\beta};\mu_{i} }\Bigg[R_{i}(\bm{a}, g, \omega_{i}|\theta_{i}) + \gamma\sum_{g'} \mathcal{T}_{g}(g'|g, \bm{a}) \Big(\sum_{t=0}^{\infty} \sum_{g''} \gamma^{t}\big(\mathcal{T}^{\bm{\pi}, \bm{\beta}, \bm{\alpha}}_{g''g'}\big)^{t}\\ 
    &\times\sum_{\bm{a}', \bm{\omega}^{k'}, \bm{\omega}' } \bm{\pi}(\bm{a}'|g'', \bm{\omega}'; \bm{\theta})R_{i}(\bm{a}', g'', \omega'_{i}|\theta_{i})  \Big)  \Bigg].
    \end{aligned}
\end{equation}

For simplicity, we will remove combine the agent's obedient selection and the signal sent by the principal for the rest of the paper unless otherwise stated, e.g., $V^{\bm{\pi}, \bm{\beta}, \bm{\alpha}; \mu_{i}}_{i} (g,  \omega^{k}_{i}|\theta_{i})=V^{\bm{\pi}, \bm{\beta}, \bm{\alpha}; \mu_{i}}_{i} (g, \omega^{k}_{i}; \omega^{k}_{i}|\theta_{i})$.
The following proposition stated without proof is an analog of Bellman's Theorem \cite{bellman1966dynamic}.
\begin{proposition}\label{prop:Bellman_theorem}
Given a game $M[\bm{\alpha}]$ with any stationary $\bm{\alpha}$, stationary $<\bm{\beta}, \bm{\pi}>$ and $\bm{\mu}$, for any $V_{i}: \mathcal{G} \times \Omega \times \Omega \mapsto \mathbb{R}$, $J_{i}: \mathcal{G} \mapsto \mathbb{R}$, $Q_{i}: \mathcal{G} \times \Omega \times \bm{\mathcal{A}} \times \Omega \mapsto \mathbb{R}$, $V_{i} = V^{\bm{\pi}, \bm{\beta}, \bm{\alpha}; \mu_{i}}_{i}: \mathcal{G} \times \Omega \times \Omega \mapsto \mathbb{R}$, $J^{\bm{\alpha}}_{i}=J^{\bm{\pi}, \bm{\beta}, \bm{\alpha}; \mu_{i}}_{i}: \mathcal{G} \mapsto \mathbb{R}$, and $Q_{i} = Q^{\bm{\pi}, \bm{\beta}, \bm{\alpha}; \mu_{i}}_{i}: \mathcal{G} \times \Omega \times \bm{\mathcal{A}} \times \Omega \mapsto \mathbb{R}$, if and only if:
\begin{equation}\label{eq:bellman_V_1}
    \begin{aligned}
    V_{i}(g, \omega_{i}; \omega^{k}_{i}|\theta_{i}) = \mathbb{E}^{\bm{\beta};\mu_{i}}\Big[\sum_{\bm{a}} \bm{\pi}(\bm{a}|g, \bm{\omega})Q_{i}  (g, \omega_{i}, \bm{a};\omega^{k}_{i}|\theta_{i})\Big],
    \end{aligned}
\end{equation}
\begin{equation}\label{eq:bellman_J_1}
    J^{\bm{\alpha}}_{i}(g|\theta_{i}) = \mathbb{E}^{\bm{\beta};\mu_{i}}\Big[\sum_{\omega_{i}, \omega^{k}_{i}}\alpha_{i}(\omega^{k}_{i}|g, \bm{\theta}) V_{i}(g, \omega_{i}; \omega^{k}_{i}|\theta_{i})\Big],
\end{equation}
\begin{equation}\label{eq:bellman_Q_1}
    \begin{aligned}
    &Q_{i}(g, \omega_{i}, \bm{a}; \omega^{k}|\theta_{i}) = R_{i}(\bm{a}, g, \omega_{i}|\theta_{i}) + \mathbb{E}^{\bm{\beta};\mu_{i}}\Big[\gamma \sum_{g', \omega'_{i}, \omega^{k'}_{i} }\mathcal{T}_{g}(g'|g, \bm{a})\alpha_{i}(\omega^{k}_{i}|g, \bm{\theta}) V_{i}(g', \omega'_{i}; \omega^{k'}_{i}|\theta_{i}) \Big].
    \end{aligned}
\end{equation}
%
%
\end{proposition}

From Proposition \ref{prop:Bellman_theorem}, we can reformulate $V^{\bm{\pi}, \bm{\beta}, \bm{\alpha}; \mu_{i}}_{i}$, $J^{\bm{\pi}, \bm{\beta}, \bm{\alpha}; \mu_{i}}_{i}$, and $Q^{\bm{\pi}, \bm{\beta}, \bm{\alpha}; \mu_{i}}_{i}$ given in (\ref{eq:state_signal_value_function})-(\ref{eq:agents_Q_function}), respectively, recursively such that (\ref{eq:bellman_V_1})-(\ref{eq:bellman_Q_1}) are satisfied.

The following proposition characterizes any PBME $<\bm{\beta}, \bm{\pi}>$ in $M[\bm{\alpha}]$ for any $\bm{\alpha}$.

\begin{proposition}\label{theorem:one_shot_deviaiton_theorem}
In any $M[\bm{\alpha}]$, a stationary strategy profile $<\bm{\beta}, \bm{\pi}>$ with $\bm{\mu}$ is a PBME if and only if, for all $i\in \mathcal{N}$, $\omega^{k}\in \Omega$ with $\alpha_{i}(\omega^{k}|g, \bm{\theta})>0$, $(\omega_{i}, a_{i})\in \Omega \times \mathcal{A}$ with $\omega_{i} = \beta_{i}( g, \omega^{k}_{i}|\theta_{i})$ and $\pi_{i}(a_{i}|g, \omega_{i})>0$, $\omega'_{i}\in \Omega$, $a'_{i}\in \mathcal{A}$,
\begin{equation}
    \begin{aligned}
    &\mathbb{E}_{\bm{a}_{-i}\sim \bm{\pi}_{-i}}\Big[ Q^{\bm{\pi}, \bm{\beta}, \bm{\alpha}; \mu_{i}}_{i}  (g, \omega_{i}, a_{i}, \bm{a}_{-i};\omega^{k}_{i}|\theta_{i})\Big] \geq   \mathbb{E}_{\bm{a}_{-i}\sim \bm{\pi}_{-i}}\Big[  Q^{\bm{\pi}, \bm{\beta}, \bm{\alpha}; \mu_{i}}_{i}  (g, \omega'_{i}, a'_{i}, \bm{a}_{-i};\omega^{k}_{i}|\theta_{i}) \Big].
    \end{aligned}
\end{equation}
\end{proposition}

Proposition \ref{theorem:one_shot_deviaiton_theorem} establishes a \textit{one-shot deviation principle}.
In particular, if it is optimal for the agent $i$ to follow the equilibrium selection rule $\beta_{i}$ and policy $\pi$ for a given observation $o_{i}$ when his behaviors for situations $o_{i}$ has transitioned from and will transition to follow the equilibrium, then it is also optimal for him to follow the equilibrium even if he has deviated in the past and will deviate in the future situations.
This implies that we can restrict attention to the characterization of the equilibrium to its robustness to one-shot deviation.
We have the following lemma.
\begin{lemma}\label{lemma:primary_equilibrium_conditions}
In a stationary PBME of a game $M[\bm{\alpha}]$, $<\bm{\beta}^{*}, \bm{\pi}^{*}>$, the following holds: for all $i\in \mathcal{N}$, $\omega^{k}\in \Omega$ with $\alpha_{i}(\omega^{k}|g, \bm{\theta})>0$, $(\omega_{i}, a_{i})\in \Omega \times \mathcal{A}$ with $\omega_{i} = \beta_{i}( g, \omega^{k}_{i},\theta_{i})$ and $\pi_{i}(a_{i}|g, \omega_{i})>0$, $\omega'_{i}\in \Omega$, $a'_{i}\in \mathcal{A}$,
\begin{equation}\label{eq:corollary_one_shot_V}
    \begin{aligned}
    &V^{\bm{\pi}, \bm{\beta}, \bm{\alpha}; \mu_{i}}_{i}(g,\omega_{i}; \omega^{k}|\theta_{i})
    \geq  \mathbb{E}^{\bm{\beta};\mu_{i}}_{\bm{a}_{-i}\sim \bm{\pi}_{-i}}\Big[  Q^{\bm{\pi}, \bm{\beta}, \bm{\alpha}; \mu_{i}}_{i}  (g, \omega'_{i}, a'_{i}, \bm{a}_{-i};\omega^{k}_{i}|\theta_{i}) \Big],
    \end{aligned}
\end{equation}
\begin{equation}\label{eq:corollary_one_shot_J}
    \begin{aligned}
    &J^{\bm{\pi}, \bm{\beta}, \bm{\alpha}; \mu_{i}}_{i}(g|\theta_{i})
    \geq \mathbb{E}^{\bm{\beta};\mu_{i}}\Big[\sum_{ \omega^{k}_{i}}\alpha_{i}(\omega^{k}|g, \bm{\theta}) V^{\bm{\pi}, \bm{\beta}, \bm{\alpha}; \mu_{i}}_{i}(g, \omega'_{i}; \omega^{k}|\theta_{i})\Big].
    \end{aligned}
\end{equation}
\end{lemma}

In Lemma \ref{lemma:primary_equilibrium_conditions}, the right hand side (RHS) of (\ref{eq:corollary_one_shot_V}) is the state-signal-action value of $<\bm{\beta}, \bm{\pi}>$ with other agents' actions $\bm{a}_{-i}$ averaged out when there are arbitrary deviations $(\omega'_{i}, a'_{i})$ from $<\bm{\beta}, \bm{\pi}>$.
The RHS of (\ref{eq:corollary_one_shot_J}) is
the expected state-signal value with the profile $<\bm{\beta}, \bm{\pi}>$ of $<\bm{\beta}, \bm{\pi}>$ with the expectation taken over the agent $i$'s selected signal when there is an arbitrary deviation $\omega'_{i}$ from $\bm{\beta}$.
Here, (\ref{eq:corollary_one_shot_V}) says that a PBME requires each agent $i$'s $V^{\bm{\pi}, \bm{\beta}, \bm{\alpha}; \mu_{i}}_{i}$ is robust to any action deviation in terms of $Q^{\bm{\pi}, \bm{\beta}, \bm{\alpha}; \mu_{i}}_{i}$, when all other agents are playing equilibrium profile $<\bm{\beta}_{-i}, \bm{\pi}_{-i}>$;
(\ref{eq:corollary_one_shot_J}) says that each agent $i$'s $J^{\bm{\pi}, \bm{\beta}, \bm{\alpha}; \mu_{i}}_{i}$ is robust to any signal selection deviation captured by $V^{\bm{\pi}, \bm{\beta}, \bm{\alpha}; \mu_{i}}_{i}$, when all other agents are playing equilibrium profile $<\bm{\beta}_{-i}, \bm{\pi}_{-i}>$.

\subsection{Characterizing OIL-P}

In a $M[\bm{\alpha}^{\text{OIL}}]$, the principal designs her signaling rule $\bm{\alpha}^{OIL}$ such that $\beta^{O}_{i}$ is obedient and $\pi^{\text{AO}}$ is admissible for each agent $i$.
Given a $\bm{\kappa}$ and an $\bm{\alpha}$, define, for any $g\in\mathcal{G}$, $\theta_{i}\in \Theta$,
\begin{equation}\label{eq:kappa_state_value}
   \begin{aligned}
   \bar{V}^{\bm{\kappa},\bm{\alpha}}_{i}(g;V_{i}|\theta_{i})\equiv&\mathbb{E}^{\mu_{i}}\Big[ \sum_{\omega^{k}_{i}}V_{i}(g,\omega^{k}_{i}|\theta_{i})\alpha_{i}(\omega^{k}_{i}|g, \bm{\theta})\Big].
   \end{aligned}
\end{equation}
Motivated by a fundamental formulation of a Nash equilibrium as a nonlinear program (see, Theorem 3.8.2 of \cite{filar1997competitive}), we obtain the following theorem that characterizes the OIL-P of the principal's information design problem.
%
%
%

\begin{theorem}\label{thm:constrained_IDP}
Suppose that the principal's goal is $\bm{\kappa}$. 
A signaling rule $\bm{\alpha}^{OIL}$ is OIL-P if and only if it induces an O-PBME $<\bm{\beta}^{O}, \bm{\pi}^{AO}>$ with $\bm{\mu}_{i}$ and the corresponding $V^{\bm{\alpha}^{OIL}}$ satisfying (\ref{eq:bellman_V_1})-(\ref{eq:bellman_Q_1}), that is the global minimum of the following constrained optimization problem with $Z^{\text{OIL-P}}(\bm{\pi}^{AO}, \bm{\beta}^{O}, \bm{V}^{\bm{\alpha}^{\text{OIL}}}; \bm{\alpha}^{\text{OIL}}, \bm{\kappa}) = 0$:
\begin{equation}\label{eq:objective_IDP_V1}
    \begin{aligned}
    &\min_{\bm{\pi}, \bm{\beta}, \bm{V}}  Z^{\text{OIL-P}}(\bm{\pi}, \bm{\beta}, \bm{V}; \bm{\alpha}^{\text{OIL}}, \bm{\kappa} )
    \equiv  \sum_{\substack{i\in \mathcal{N},  \\g,\omega_{i}}} \bar{V}^{\bm{\kappa},\bm{\alpha}^{OIL}}_{i}(g;V_{i}|\theta_{i})
    - \mathbb{E}^{\omega^{k}_{i}\sim\alpha_{i}^{\text{OIL}}}_{\bm{a}\sim\bm{\pi}}\Big[ Q^{\bm{\pi}, \bm{\beta}, \bm{\alpha}^{\text{OIL}}}_{i}  (g, \omega_{i},  \bm{a};\omega^{k}_{i}|\theta_{i})\Big],
    \end{aligned}
\end{equation}
such that, for all $i\in\mathcal{N}$, $g\in\mathcal{G}$, $\omega^{k}_{i}\in \Omega$ with $\alpha^{OIL}_{i}(\omega^{k}_{i}|g, \theta_{i})>0$, $\theta_{i}\in \Theta$, $a_{i}\in\mathcal{A}$, any $\beta'_{i}$, any $\pi'_{i}$
\begin{align}
    \begin{split}\label{eq:IPD_constraint_1_1}
        V&{}_{i}(g,\omega^{k}_{i}|\theta_{i})
        \geq \mathbb{E}_{\bm{a}_{-i}\sim \bm{\pi}_{-i}}\Big[ Q^{\pi'_{i},\bm{\pi}_{-i}, \bm{\beta}, \bm{\alpha}^{\text{OIL}}}_{i}  (g, \omega^{k}_{i}, a_{i}, \bm{a}_{-i}|\theta_{i})\Big],
    \end{split}\\
    \begin{split}\label{eq:IPD_constraint_1_2}
        J&{}_{i}^{\bm{\alpha}^{\text{OIL}} }(g|\theta_{i})
        \geq  \mathbb{E}^{\bm{\beta} }\Big[\sum_{ \omega^{k}_{i}}\alpha^{\text{OIL}}_{i}(\omega^{k}_{i}|g, \bm{\theta}) V_{i}(g, \beta'_{i}(g, \omega^{k}_{i},\theta_{i}) ; \omega^{k}_{i}|\theta_{i})\Big],
    \end{split}\\
    \begin{split}\label{eq:idp_admissible_constraint}
        \bar{V}&{}^{\bm{\kappa}, \bm{\alpha}^{\text{OIL}}}_{i}(g;V_{i}|\theta_{i})
        = \mathbb{E}^{ \alpha^{\text{OIL}}_{i}}_{ \bm{\pi} }\Big[ Q^{\bm{\pi}, \bm{\beta}, \bm{\alpha}^{\text{OIL}}}_{i}  (g, \omega^{k}_{i}, \bm{a}|\theta_{i})\Big],
    \end{split}
\end{align}
where $J_{i}^{\bm{\alpha}^{\text{OIL}}}$ and $Q_{i}^{\bm{\pi}, \bm{\beta}, \bm{\alpha}^{\text{OIL}}}$ are constructed according to (\ref{eq:bellman_J_1}) and (\ref{eq:bellman_Q_1}) in terms of $V_{i}$.
%
%
\end{theorem}
%


In Theorem \ref{thm:constrained_IDP}, the optimization problem has three decision variables, $\bm{\pi}$, $\bm{\beta}$, and $\bm{V}$. 
It is straightforward to see that the objective function in (\ref{eq:objective_IDP_V1}), $Z^{\text{OIL-P}}(\bm{\pi}^{AO}, \bm{\beta}^{O},$  $\bm{V}^{O}; \bm{\alpha}^{\text{OIL}}, \bm{\kappa}) = 0$, at an O-PBME, given a principal's goal $\bm{\kappa}$.
Hence, in an O-PBME, only the three constraints remain in the constrained in the optimization problem.
The constraint (\ref{eq:IPD_constraint_1_1}) requires that given the obedient $\bm{\beta}^{O}$, each agent $i$ has no incentive to deviate from an equilibrium policy profile $\bm{\pi}$ by any arbitrary deviation $a_{i}$.
In other words, the constraint (\ref{eq:IPD_constraint_1_1}) guarantees that the signaling rule $\bm{\alpha}^{\text{OIL}}$ and the obedient $\bm{\beta}^{O}$ lead the agents to an PBME.
%
%
The constraint (\ref{eq:IPD_constraint_1_2}) requires that for any equilibrium policy profile, the obedient $\bm{\beta}^{O}$ is always preferred by the agents than any other selection rule.
Finally, the constraint (\ref{eq:idp_admissible_constraint}) requires that the equilibrium policy $\bm{\pi}^{AO}$ in the PBME with obedient $\bm{\beta}^{O}$ is admissible given $\bm{\kappa}$.
Together, the constraints (\ref{eq:IPD_constraint_1_1}), (\ref{eq:IPD_constraint_1_2}), and (\ref{eq:idp_admissible_constraint}) guarantee that obedient $\bm{\beta}^{O}$ and the corresponding policy profile $\bm{\pi}^{AO}$ constitute an admissible O-PBME, such that the design of $\bm{\alpha}^{O}$ incentivizes the agents to behave according to $<\bm{\beta}^{O}, \bm{\pi}^{AO}>$ instead of choosing any other non-obedient equilibrium $<\bm{\beta}^{*}, \bm{\pi}^{*}>$ or any arbitrary deviations.
For a signaling rule $\bm{\alpha}$ and a profile $<\bm{\beta},\bm{\pi}>$, we define the \textit{occupancy measure}, denoted as $\bm{\rho}_{\bm{\pi}}^{\bm{\alpha}, \bm{\beta}}$, of the $\bm{\alpha}$-augmented game $M[\bm{\alpha}]$ as follows: for any $g\in\mathcal{G}$, $\bm{a}\in\bm{\mathcal{A}}$, $\bm{\omega}\in \Omega^{n}$, $\bm{\omega}^{k}\in \Omega^{n}$, $\bm{\theta}\in \bm{\Theta}$,
\begin{equation}\label{eq:occupancy_measure_1}
    \begin{aligned}
    \bm{\rho}&{}_{\bm{\pi}}^{\bm{\alpha}, \bm{\beta}}(g, \bm{a}, \bm{\omega},  \bm{\omega}^{k}|\bm{\theta})
    \equiv \bm{\alpha}(\bm{\omega}^{k}|g, \bm{\theta})\bm{\rho}^{\bm{\beta}}_{\bm{\pi}}(g, \bm{a}, \bm{\omega}|\bm{\omega}^{k}, \bm{\theta}), 
    \end{aligned}
\end{equation}
%
%
where 
\begin{equation}\label{eq:short_occupancy}
    \begin{aligned}
    &\bm{\rho}^{\bm{\beta}}_{\bm{\pi}}(g, \bm{a}, \bm{\omega}|\bm{\omega}^{k}, \bm{\theta})
    %
    %
    \equiv \bm{\pi}(\bm{a}|g,\bm{\omega}, \bm{\theta}) \times\sum_{t=0}^{\infty}\gamma^{t} P(g_{t}=g, \bm{\beta}(g, \bm{\omega}^{k}, \bm{\theta}) = \bm{\omega}|\bm{\omega}^{k}_{t} = \bm{\omega}^{k}; \bm{\beta}, \bm{\pi}),
    \end{aligned}
\end{equation}
is the \textit{signal-conditioned occupancy measure} 
%
%
For notational compactness, we let $\bm{\rho}_{\bm{\pi}}^{\bm{\alpha}, \bm{\beta}}(g, \bm{a}, \bm{\omega}^{k}|\bm{\theta}) = \bm{\rho}_{\bm{\pi}}^{\bm{\alpha}, \bm{\beta}}(g, \bm{a}, \bm{\omega}^{k},  \bm{\omega}^{k}|\bm{\theta})$, unless otherwise stated.
Given the belief system $\bm{\mu}$, the occupancy measure perceived by each agent $i$ is given as follows:
\begin{equation}
    \begin{aligned}
    \rho&{}_{[i],\bm{\pi}}^{\bm{\alpha}, \bm{\beta};\mu_{i}}(g, \bm{a}, \omega_{i},  \omega^{k}_{i}|\theta_{i})\equiv \mathbb{E}^{\mu_{i}}\Big[\bm{\rho}_{\bm{\pi}}^{\bm{\alpha}, \bm{\beta}}(g, \bm{a}, \bm{\omega},  \bm{\omega}^{k}|\bm{\theta}) \Big],
    \end{aligned}
\end{equation}
with $\rho_{[i],\bm{\pi}}^{\bm{\alpha}, \bm{\beta};\mu_{i}}(g, \bm{a}, \omega^{k}_{i}|\theta_{i}) = \rho_{\bm{\pi},[i]}^{\bm{\alpha}, \bm{\beta};\mu_{i}}(g, \bm{a}, \omega^{k}_{i},  \omega^{k}_{i}|\theta_{i})$.
Similar, given the principal's goal $\bm{\kappa}$, we can define the occupancy measure with respect to $\bm{\kappa}$: $\bm{\rho}^{\bm{\kappa}}(g,\bm{a}|\bm{\theta})\equiv \sum_{t=0}^{\infty}P^{\bm{\kappa}}(g_{t} = g, \bm{a}_{t} = \bm{a}|\bm{\theta})$, where $P^{\bm{\kappa}}$ represents the probability given $\bm{\kappa}$ and the transition of the global state.
Similarly, we define the occupancy measure associated with the principal's goal $\bm{\kappa}$ as 
\begin{equation}\label{eq:occupancy_kappa}
    \bm{\rho}^{\bm{\kappa}}(g,\bm{a} |\bm{\theta})\equiv \bm{\kappa}(\bm{a}|g,\bm{\theta})\sum_{t=0}^{\infty}P(g_{t}=g|\bm{\theta};\bm{\kappa}).
\end{equation}
We extends the basic result known as \textit{Bellman flow constraints} of Markov decision process (see, e.g., \cite{puterman2014markov,syed2008apprenticeship,ho2016generative}) to the $M[\bm{\alpha}]$ and define the following set of occupancy measures:
\begin{equation}
    \begin{aligned}
    \mathcal{D}[\bm{\alpha};\bm{\theta}]\equiv \Bigg\{\bm{\rho}: \bm{\rho}\geq 0, &\text{ and } 
    \sum_{\bm{a}, \bm{\omega}}\bm{\rho}\big(g, \bm{a},\bm{\omega}, \bm{\omega}^{k}\big| \bm{\theta}\big) \\
    &= \bm{\alpha}(\bm{\omega}^{k}|g, \bm{\theta})\Big( d_{g}(g) 
    + \gamma\sum_{g',\bm{a}', \bm{\omega}' } \bm{\rho}\big(g', \bm{a}', \bm{\omega}',\bm{\omega}^{k}\big| \bm{\theta}\big) \mathcal{T}_{g}(g|g', \bm{a}') \Big)\Bigg\}.
    \end{aligned}
\end{equation}
%
%
%

We have the following corollary that characterizes the results in Theorem \ref{thm:constrained_IDP} in terms of the occupancy measure.


\begin{corollary}\label{corollary:occupancy_linear}
Fix a $\bm{\kappa}$.
The signaling rule $\bm{\alpha}^{*}$ is OIL-P if and only if there exists a $\bm{\rho}^{*}\in \mathcal{D}[\bm{\alpha}^{*};\bm{\theta}]$ that solves the following constrained optimization problem:
\begin{equation}\label{eq:occupancy_linear_1}
    \max\limits_{\bm{\rho}\in\mathcal{D}[\bm{\alpha}^{*};\bm{\theta} ]} \sum_{g, \bm{a}, \bm{\omega}^{k} }\sum_{i\in\mathcal{N}}R_{i}(g, \bm{a}, \omega^{k}_{i}|\theta_{i})
    \bm{\rho}(g, \bm{a},  \bm{\omega}^{k}|\bm{\theta}),
    %
\end{equation}
such that, for any $g\in\mathcal{G}$, $\bm{\omega}^{k}\in \Omega^{n}$, $\bm{a}\in\bm{\mathcal{A}}$, $\bm{\theta}\in \bm{\Theta}$, 
\begin{equation}\label{eq:occupancy_obedient}
    \begin{split}
         &\frac{\sum_{\bm{a}'}\bm{\rho}\big(g, \bm{a}',\bm{\omega}^{k}\big| \bm{\theta}\big)  }{\sum_{\bm{a}', \bm{\omega}, }\bm{\rho}\big(g, \bm{a}', \bm{\omega}, \bm{\omega}^{k}\big| \bm{\theta}\big) } = 1,
    \end{split}
\end{equation}
\begin{equation}\label{eq:occupancy_admissible_1}
    \sum_{\bm{\omega}^{k}}\bm{\rho}(g,\bm{a}, \bm{\omega}^{k}|\bm{\theta})\bm{\alpha}(\bm{\omega}^{k}|g,\bm{\theta}) =\bm{\rho}^{\bm{\kappa}}(g,\bm{a} |\bm{\theta}).
\end{equation}
For each $\rho^{*}$ that solves (\ref{eq:occupancy_linear_1})-(\ref{eq:occupancy_admissible_1}), the admissible O-PBME policy profile $\bm{\pi}^{AO}_{\rho^{*}}$ is uniquely given as, for any $g\in\mathcal{G}$, $\bm{\omega}^{k}\in \Omega^{n}$, $\bm{a}\in\bm{\mathcal{A}}$, $\bm{\theta}\in \bm{\Theta}$,
\begin{equation}\label{eq:occupancy_measure_policy}
    \bm{\pi}^{AO}_{\rho^{*}}(\bm{a}|g, \bm{\omega}^{k}, \bm{\theta})=\frac{\bm{\rho}^{*}(g, \bm{a}, \bm{\omega}^{k}|\bm{\theta})}{\sum_{\bm{a}'}  \bm{\rho}^{*}(g, \bm{a}', \bm{\omega}^{k}|\bm{\theta})}.
\end{equation}
\end{corollary}
%


Corollary \ref{corollary:occupancy_linear} extends the basic linear programming formulation of Markov decision process in terms of the occupancy measure (see, e.g., \cite{puterman2014markov,syed2008apprenticeship,ho2016generative}) to the game $M[\bm{\alpha}]$ in O-PBME and formulate the constrained optimization problem (\ref{eq:objective_IDP_V1})-(\ref{eq:idp_admissible_constraint}) in Theorem \ref{thm:constrained_IDP} as an occupancy measure selection problem.
Choosing an occupancy measure from the set $\mathcal{D}^{O}[\bm{\alpha}^{*};\bm{\theta}]$ captures the \textit{Bellman flow constraints} \cite{syed2008apprenticeship} of the game $M[\bm{\alpha}^{*}]$ when the agents use obedient $\bm{\beta}^{O}$.
The constraint (\ref{eq:occupancy_obedient}) requires the feasible occupancy measures are those with obedient $\bm{\beta}^{O}$, i.e., the probability of $\bm{\beta}(g,\bm{\omega}^{k},\bm{\theta}) = \bm{\omega}^{k}$ is $1$.
The second constraint (\ref{eq:occupancy_admissible_1}) guarantees the admissibility of the optimal policy profile associated with the optimal occupancy measure $\bm{\rho}^{*}$.
Here, (\ref{eq:occupancy_measure_policy}) is from a basic result that for each occupancy measure $\bm{\rho} \in \mathcal{D}^{O}[\bm{\alpha}^{*};\bm{\theta}]$, there is a unique policy profile that can be constructed in terms of occupancy measure.

Next, we extend the occupancy measure and define the \textit{$t$-sequential occupancy measure} perceived by each agent $i$, denoted as $\lambda^{\bm{\alpha}, \bm{\beta}}_{\bm{\pi},[i],t}$, which is the distribution of sequences of state-action-signals of length $t$ that the agents encounters when using the equilibrium profile $<\bm{\beta},\bm{\pi}>$, given a signaling rule $\bm{\alpha}$.
Define \textit{trajectory} of length $t-\tau+1$ $x^{(t)}_{i;\tau}\equiv \{x_{i,t}, x_{i,t-1}, \dots, x_{i,\tau}\}$ and $\bm{x}^{(t)}_{\tau}\equiv\{\bm{x}_{t}, \bm{x}_{t-1},\dots, \bm{x}_{\tau}\}$, respectively, for the term $x$ and the joint $\bm{x}=\{x_{i}\}_{i\in\mathcal{N}}$ of $x$, for $x=g, a, \omega, \omega^{k}$ with $\omega^{k;(t)}_{i;\tau}$ and $\bm{\omega}^{k;(t)}_{\tau}$, $i\in \mathcal{N}$.
Define the \textit{sequence} of length $t$ as
\begin{equation*}
    \begin{aligned}
    &h_{\tau:t}[g^{(t)}_{\tau}, \bm{\omega}^{k;(t)}_{\tau}, \bm{\omega}^{(t)}_{\tau}, \bm{a}^{(t)}_{\tau}]\equiv \{(g_{\tau}, \bm{\omega}^{k}_{\tau}),
    (\bm{\omega}_{\tau},
    \bm{a}_{\tau},g_{\tau+1}, 
    \bm{\omega}^{k}_{\tau+1}),
    \dots,  (\bm{\omega}_{\tau+t-1}, \bm{a}_{\tau+t-1}, g_{\tau+t}, \bm{\omega}^{k}_{\tau+t})\}\in H_{t}, 
    \end{aligned}
\end{equation*}
with $h_{\tau:t}[g^{(t)}_{\tau}, \bm{\omega}^{k;(t)}_{\tau},\bm{a}^{(t)}_{\tau}] = h_{\tau:t}[g^{(t)}_{\tau}, \bm{\omega}^{k;(t)}_{\tau}, \bm{\omega}^{k;(t)}_{\tau}, \bm{a}^{(t)}_{\tau}]$, where 
where $H_{t}\equiv \mathcal{G}\times \Omega^{n} \times (\Omega^{n} \times \bm{\mathcal{A}} \times \mathcal{G}\times \Omega^{n})^{t}$.
Similar, we define $\ell_{\tau:t}[g^{(t)}_{\tau}, \bm{\omega}^{k;(t)}_{\tau}, \bm{\omega}^{(t)}_{\tau}]\in H^{\backslash \bm{a}}_{t}$ as $h_{t}[g^{(t)}, \bm{\omega}^{k;(t)}, \bm{\omega}^{(t)}, \bm{a}^{(t)}]$ without trajectory of actions $\bm{a}^{(t)}$, where $H^{\backslash \bm{a}}_{t}\equiv \mathcal{G}\times \Omega^{n} \times (\Omega^{n} \times \mathcal{G}\times \Omega^{n})^{t}$.
For notational compactness, we simplify the sequences by only showing $h_{t}$ and $\ell_{t}$ without trajectories or show some specific trajectories or elements for the purpose of highlight (e.g., a realized sequence $h_{t}$ can be written as $h_{t}[\bm{a}^{t}]$ or $h_{t}[\bm{a}_{t}, \bm{\omega}^{k}_{t}]$), unless otherwise stated.
We will write $\ell_{\tau:\tau+t}$ or $\subset h_{\tau:\tau+t}$ if $\ell_{\tau:\tau+t}\subset h_{\tau:\tau+t}[\bm{a}^{t}]$ if $\ell_{\tau:\tau+t}$ is $h_{\tau:\tau+t}[\bm{a}^{t}]$ without $\bm{a}^{t}$.
When $\tau=0$, the time index of $\tau$ is ignored in the notations of the trajectory and the sequences, e.g.,  $x^{(t)} = x^{(t)}_{0}$ and $h_{t} = h_{0:t}$.

Formally, the $t$-sequential occupancy measure is defined as, for any $h_{t}\in H_{t}$, $\bm{\theta}\in \bm{\Theta}$,
\vspace{-0.5\baselineskip}
\begin{equation}\label{eq:sequential_occupancy}
    \begin{aligned}
    \lambda^{\bm{\alpha},\bm{\beta}}_{\bm{\pi},t}(h_{t}|\bm{\theta})\equiv \sum_{\tau=0}^{\infty} P^{\bm{\alpha},\bm{\beta}}_{\bm{\pi}}\big(h_{\tau:\tau+t} = h_{t} |\bm{\theta}\big).
    \end{aligned}
\end{equation}
Similarly, given any $g^{(t)}\subset \ell_{t}\subset h_{t}[\bm{a}^{(t)}]\in H_{t}$, we have 
$\bm{\bar}{\lambda}^{\bm{\alpha},\bm{\beta}}_{\bm{\pi},t}(\ell_{t}|\bm{\theta})\equiv \sum_{\bm{a}'} \lambda^{\bm{\alpha},\bm{\beta}}_{\bm{\pi},t}(h_{t}[\bm{a}']|\bm{\theta})$ 
%
%
For any $\tilde{h}_{t}=\{\tilde{h}_{i,t}\}_{i\in\mathcal{N}}=\{(\tilde{g}_{\tau}, \tilde{\omega}^{k}_{i,\tau},$  $\tilde{\omega}_{i,\tau}, \tilde{\bm{a}}_{\tau}\}_{i\in\mathcal{N},\tau=0}^{t}\in H_{t}$, we denote, for any $\tau\geq0$, $t\geq 0$, 
$$
\begin{aligned}
&R^{(\tau+t)}_{i;\tau}(\tilde{h}_{i,t}|\theta_{i})\equiv  \sum_{s=0}^{t}\gamma^{\tau+s} R_{i}(g_{\tau+s} = \tilde{g}_{s}, \bm{a}_{\tau+s}=\tilde{\bm{a}}_{s}, \omega_{i,\tau+s} = \tilde{\omega}_{i,s}|\theta_{i}).
\end{aligned}
$$
%
%
Then, the following holds:
$$
\begin{aligned}
    \mathbb{E}^{\bm{\alpha},\bm{\beta}}_{\bm{\pi}}\Big[ R_{i}(g', a', \omega'_{i}|\theta_{i})\Big] &= \sum\limits_{\substack{h_{t}\in H_{t},\\i\in\mathcal{N}}} R^{(t)}_{i}(h_{i,t}|\theta_{i})\lambda^{\bm{\alpha},\bm{\beta}}_{\bm{\pi},t}(h_{t}|\bm{\theta}).\\
    &= \sum_{g, \bm{a}, \bm{\omega}^{k} }\sum_{i\in\mathcal{N}}R_{i}(g, \bm{a}, \omega^{k}_{i}|\theta_{i})
    \bm{\rho}^{\bm{\beta},\bm{\beta}}_{\bm{\pi}}(g, \bm{a},  \bm{\omega}^{k}|\bm{\theta}).
\end{aligned}
$$
Given the belief $\mu_{i}$, we let $\lambda^{\bm{\alpha},\bm{\beta};\mu_{i}}_{[i],\bm{\pi},t}(h_{i,t}|\theta_{i})$ denote the $t$-sequential occupancy measure perceived by each agent $i$: $\lambda^{\bm{\alpha},\bm{\beta};\mu_{i}}_{[i],\bm{\pi},t}(h_{i,t}|\theta_{i}) = \mathbb{E}^{\mu_{i}}\Big[\sum\limits_{\bm{\omega}^{k;(t)}_{-i}, \bm{\omega}^{(t)}_{-i} } \lambda^{\bm{\alpha},\bm{\beta}}_{\bm{\pi},t}(h_{t}|\bm{\theta})\Big]$. 
Analogously, we can define $\bm{\bar}{\lambda}^{\bm{\alpha},\bm{\beta};\mu_{i}}_{[i],\bm{\pi},t}$.
%
%
%

Let $h_{i,t}$ denote any sequence of length $t$ (perceived by agent $i$), $t>0$.
For any $t'$ with $0\leq t' < t$, we write $h_{i,t}=h_{i,t'} \oplus h_{i,t'+1:t}$ such that $h_{i,t'}$ is the first $t'$ components of $h_{i,t}$ and $h_{i,t'+1:t}$ is the last $t-t'$ sequence of $h_{i,t}$; $\oplus$ is not symmetric in general, i.e., $h_{i,t}+h'_{i,t} \neq h'_{i,t} + h_{i,t}$.
Given any two sequences $h_{i,t}$ and $h'_{i,t'}$ of lengths $t$ and $t'$ and any sequence $\ell_{i,t''}$ of length $t''$, the transition functions of the sequences can be formulated as follows:
%
%
\begin{align*}
    \begin{split}
        \mathcal{T}{}^{\bm{\pi}, \bm{\beta},\bm{\alpha}}_{h_{i,t'}, h_{i,t}} = \mathcal{T}_{h}(h_{i,t'}| h_{i,t}; \theta_{i}) 
    \equiv \frac{\lambda^{\bm{\alpha}, \bm{\beta};\mu_{i}}_{[i],\bm{\pi},t+t'}(h_{i,t} \oplus h_{i,t'}|\theta_{i})  }{\lambda^{\bm{\alpha}, \bm{\beta};\mu_{i}}_{[i],\bm{\pi},t}(h_{i,t}|\theta_{i})},
    \end{split}\\
    \begin{split}
        &\mathcal{T}{}^{\bm{\pi}, \bm{\beta},\bm{\alpha}}_{\ell_{i,t''}, h_{i,t}} = \mathcal{T}_{\ell h}(\ell_{i,t''}|h_{i,t})\equiv \frac{\sum_{\bm{a}^{(t'')}}\lambda^{\bm{\alpha}, \bm{\beta};\mu_{i}}_{[i],\bm{\pi},t+t''}(h_{i,t} \oplus \ell_{i,t''}\cup\{\bm{a}^{(t'')}\}|\theta_{i})  }{\lambda^{\bm{\alpha}, \bm{\beta};\mu_{i}}_{[i],\bm{\pi},t}(h_{i,t}|\theta_{i})}.
    \end{split}
\end{align*}
%
Hence, $\mathcal{T}^{\bm{\pi}, \bm{\beta},\bm{\alpha}}_{h_{i,t'}, h_{i,t}}$ and $\mathcal{T}{}^{\bm{\pi}, \bm{\beta},\bm{\alpha}}_{\ell_{i,t''}, h_{i,t}}$, respectively, give the probability of the next sequences $h_{i,t'}$ of length $t'$ and $\ell_{i,t''}$ of length $t''$, given the current sequence $h_{i,t}$.
The sequence transition function, $\mathcal{T}_{\ell'_{i,t'}, \ell_{i,t}}$, associated with $\bar{\lambda}^{\bm{\alpha},\bm{\beta}}_{\bm{\pi},[i],t}$ can be defined in the same way, for any two sequences $\ell'_{i,t'}\in H_{i,t'}^{\backslash \bm{a}}$ and $\ell_{i,t} \in H_{i,t}^{\backslash \bm{a}}$, $t,t'\geq0$.

Given any $h_{t}=\{h_{i,t}\}_{i\in\mathcal{N}}\in H_{t}$, $g^{t}\subset \ell_{t}\subset h_{t}$, for any $t\geq0$, we define the following value functions, $Q^{\bm{\pi}, \bm{\beta}, \bm{\alpha};\mu_{i}}_{i|t,t'}$ and $V^{\bm{\pi}, \bm{\beta}, \bm{\alpha};\mu_{i}}_{i|t,t'}$, with a slight abuse of notation: for any $t,t',t''\geq0$, for all $i\in\mathcal{N}$,
\begin{align}
    \begin{split}\label{eq:sequential_occupancy_Q}
        &Q{}^{\bm{\pi}, \bm{\beta}, \bm{\alpha};\mu_{i}}_{i|t,t'}(h_{i,t}|\theta_{i})\equiv R^{(t)}_{i}(h_{i,t}|\theta_{i})+
    \sum_{\ell'_{i,t'}}\mathcal{T}^{\bm{\pi},\bm{\beta},\bm{\alpha}}_{\ell'_{i,t'},h_{i,t}} V^{\bm{\pi}, \bm{\beta}, \bm{\alpha};\mu_{i}}_{i|t',t''}(\ell'_{i,t'}|\theta_{i}),
    \end{split}\\
    \begin{split}\label{eq:sequential_occupancy_V}
        &V{}^{\bm{\pi}, \bm{\beta}, \bm{\alpha};\mu_{i}}_{i|t,t'}(\ell_{i,t}|\theta_{i})\equiv \mathbb{E}^{\mu_{i}}_{\bm{\pi}}\Big[ Q^{\bm{\pi}, \bm{\beta}, \bm{\alpha};\mu_{i}}_{i|t,t'}(h_{i,t}[\bm{a}^{t}]|\theta_{i})\Big]\Big|_{\ell_{i,t}\subset h_{i,t}}.
    \end{split}
\end{align}
We refer to (\ref{eq:sequential_occupancy_Q}) and (\ref{eq:sequential_occupancy_V}) as the \textit{extended Bellman equations}.

The following lemma shows an asymptotic relationship between the regular and the sequential occupancy measures.
\begin{lemma}\label{lemma:occupancy_relationship}
Given any $<\bm{\beta}, \bm{\pi}>$ and $\bm{\alpha}$, the following holds: for all $i\in\mathcal{N}$,
\begin{equation}
    \begin{split}
        &\lim{}_{t-t'\rightarrow \infty} \sum_{h_{i,t}} Q^{\bm{\pi}, \bm{\beta}, \bm{\alpha};\mu_{i}}_{i|t,t'}(h_{i,t}|\theta_{i})\lambda^{\bm{\alpha},\bm{\beta};\mu_{i}}_{[i],\bm{\pi}, t}(h_{i,t}|\theta_{i})
        = \lim{}_{t-t'\rightarrow \infty}\sum_{\ell_{i,t}} V^{\bm{\pi}, \bm{\beta}, \bm{\alpha};\mu_{i}}_{i|t,t'}(\ell_{i,t}|\theta_{i})\bar{\lambda}^{\bm{\alpha},\bm{\beta};\mu_{i}}_{[i],\bm{\pi}, t}(\ell_{i,t}|\theta_{i})\\
        =&\sum_{g, \bm{a}, \omega_{i},\omega^{k} } R_{i}(g, \bm{a}, \omega_{i}|\theta_{i})
    \rho^{\bm{\alpha},\bm{\beta};\mu_{i}}_{[i],\bm{\pi}}(g, \bm{a}, \omega_{i}, \omega^{k}_{i}|\theta_{i}).
    \end{split}
\end{equation}
\end{lemma}


%
The following proposition re-write the constraints (\ref{eq:IPD_constraint_1_1}) and (\ref{eq:IPD_constraint_1_2}) in Theorem \ref{thm:constrained_IDP} in terms of $Q^{\bm{\pi}, \bm{\beta}, \bm{\alpha};\mu_{i}}_{i|t,t'}$ and $V^{\bm{\pi}, \bm{\beta}, \bm{\alpha};\mu_{i}}_{i|t,t'}$.
\begin{proposition}\label{prop:reformulate_OIL_constraints}
Fix a $\bm{\kappa}$.
Let $h_{i,t}\equiv \{g_{0}, \bm{a}_{0}, \omega_{i,0}, \omega^{k}_{i,0}\}\oplus h_{i,1:t}\in H_{i,t}$ and $\ell_{i,t} \subset h_{i,t}$, for any $t\geq 0$.
%
The signaling rule $\bm{\alpha}^{\text{OIL}}$ is OIL-P by $<\bm{\beta}^{O}, \bm{\pi}^{AO}>$ if and only if $\bm{\pi}^{AO}$ is admissible and, for any $\hat{a}^{(t)}_{i}\in \mathcal{A}^{t}$, $\hat{\omega}^{(t)}_{i}\in \Omega^{t}$, $t,t'\geq 0$,
\begin{equation}\label{eq:sequential_constraint_1}
    \begin{split}
        V_{i}^{\bm{\pi}^{AO},\bm{\beta}^{O}, \bm{\alpha}^{\text{OIL}}}&(g_{0},\omega^{k}_{i,0}|\theta_{i})\geq  \mathbb{E}^{\mu_{i}}_{\bm{\pi}^{AO}_{-i}}\Big[ Q^{\pi_{i},\bm{\pi}^{AO}_{-i},  \bm{\beta}^{O}, \bm{\alpha};\mu_{i}}_{i|t,t'}(h_{i,t}[\hat{a}^{(t)}_{i},\bm{a}^{t}_{-i}]|\theta_{i})\Big],
    \end{split}
\end{equation}
\begin{equation}\label{eq:sequential_constraint_2}
    \begin{split}
         J_{i}^{\bm{\pi}^{AO},\bm{\beta}^{O}, \bm{\alpha}^{\text{OIL}}}&(g_{0}|\theta_{i})
        \geq \mathbb{E}^{\bm{\alpha}^{\text{OIL}}} \Big[
        V^{\bm{\pi}^{AO}, \beta_{i}, \bm{\beta}^{O}_{-i}, \bm{\alpha};\mu_{i}}_{i|t,t'}(\ell_{i,t}[\hat{\omega}^{(t)}_{i}, \omega^{k;(t)}_{i}]|\theta_{i})\Big],
    \end{split}
\end{equation}
%
%
where $V_{i}^{\bm{\pi}^{AO},\bm{\beta}^{O}, \bm{\alpha}^{\text{OIL}}}$, $J_{i}^{\bm{\pi}^{AO},\bm{\beta}^{O}, \bm{\alpha}^{\text{OIL}}}$ satisfy (\ref{eq:bellman_V_1})-(\ref{eq:bellman_Q_1}) in Proposition \ref{prop:Bellman_theorem}, $Q{}^{\bm{\pi}, \bm{\beta}, \bm{\alpha};\mu_{i}}_{i|t,t'}$ and $V{}^{\bm{\pi}, \bm{\beta}, \bm{\alpha};\mu_{i}}_{i|t,t'}$ are given in (\ref{eq:sequential_occupancy_Q}) and (\ref{eq:sequential_occupancy_V}), respectively.
%
%
\end{proposition}
%


Given the principal's goal $\bm{\kappa}$, the restrictions imposed by $\bm{\alpha}^{\text{OIL}}$ on $\bm{\beta}$ and $\bm{\pi}$ through (\ref{eq:IPD_constraint_1_1})-(\ref{eq:IPD_constraint_1_2}) are equivalent to (\ref{eq:sequential_constraint_1})-(\ref{eq:sequential_constraint_2}) in Proposition \ref{prop:reformulate_OIL_constraints}, respectively.
Theorem \ref{thm:constrained_IDP} shows that an OIL-P signaling rule $\bm{\alpha}^{\text{OIL}}$ leads to an O-PBME solves the constrained problem with $Z^{\text{OIL-P}}(\bm{\pi}^{AO}, \bm{\beta}^{O},$ $\bm{V}^{\bm{\alpha}^{\text{OIL}}}; \bm{\alpha}^{\text{OIL}}, \bm{\kappa}) = 0$.
We introduce the \textit{slack variables}, $\delta_{\bm{\pi}^{*};t,t'}^{\bm{\beta}^{*}, \bm{\alpha} }[\pi_{i}]\geq 0$ and $\zeta_{t,t'}^{ \bm{\alpha}^{OIL}}[\beta_{i}]\geq0$, to make the inequality constraints (\ref{eq:sequential_constraint_1}) and (\ref{eq:sequential_constraint_2}) be equality, for each deviation $\pi_{i}$ and $\beta_{i}$, respectively.
Let $L^{\bm{\beta}^{*}, \bm{\alpha}}_{\bm{\pi}^{*}}(\bm{\Lambda}_{t,t'},  \bm{\Xi}_{t,t'}, \bm{V}^{*}; \bm{\delta}_{t,t'}^{\bm{\alpha}^{OIL}}, \bm{\zeta}_{t,t'}^{\bm{\alpha}^{OIL}}|\bm{\theta})$ denote the \textit{Lagrangian} of the problem (\ref{eq:objective_IDP_V1}) with constraints (\ref{eq:sequential_constraint_1}), (\ref{eq:sequential_constraint_2}), and (\ref{eq:idp_admissible_constraint}), where $\bm{\Lambda}_{t,t'}\equiv\{\Lambda^{\pi_{i}}_{i;t,t'}\}_{i\in\mathcal{N},\pi_{i}},  \bm{\Xi}\equiv\{\Xi^{\beta_{i}}_{i}\}_{i\in\mathcal{N},\beta_{i}}$ are the dual variables associated with the constraints (\ref{eq:sequential_constraint_1}) and (\ref{eq:sequential_constraint_2}), for all possible deviations $\bm{\pi}=\{\pi_{i}\}_{i\in \mathcal{N}}$ and $\bm{\beta}=\{\beta_{i}\}_{i\in \mathcal{N}}$ respectively; and 
$\bm{\delta}_{t,t'}^{\bm{\alpha}^{OIL}}\equiv\{\delta_{t,t'}^{\bm{\alpha}^{OIL}}[\pi_{i}] \}_{\pi_{i}, i\in\mathcal{N}}$ and $\bm{\zeta}_{t,t,}^{\bm{\alpha}^{OIL}}\equiv\{\zeta_{t,t'}^{\bm{\alpha}^{OIL}}[\beta_{i}] \}_{\beta_{i}, i\in\mathcal{N}}$.
Hence, the Lagrangian of the problem in Theorem \ref{thm:constrained_IDP} at an admissible O-PBME $<\bm{\beta}^{O}, \bm{\pi}^{AO}>$ takes the following form:
for any $t,t'\geq 0$,
\begin{equation}\label{eq:Lagrangian_1}
    \begin{split}
        &L^{\bm{\beta}^{O}, \bm{\alpha} }_{\bm{\pi}^{AO}}(\bm{\Lambda}_{t,t'}, \bm{\Xi}_{t,t'}, \bm{V}^{\bm{\pi}^{AO}, \bm{\beta}^{O}, \bm{\alpha}} ;\bm{\delta}_{t,t'}^{\bm{\alpha} }, \bm{\zeta}_{t,t'}^{\bm{\alpha} }|\bm{\theta})
        \equiv \sum_{i\in \mathcal{N}, \pi_{i}} \Lambda^{\pi_{i}}_{i;t,t'}\Bigg(\delta_{t,t'}^{ \bm{\alpha}^{OIL}}[\pi_{i}]-V_{i}^{\bm{\pi}^{AO},\bm{\beta}_{t,t'}^{O}, \bm{\alpha}}(g_{0},\omega^{k}_{i,0}|\theta_{i})\\
        +&\mathbb{E}^{\mu_{i}}_{\bm{\pi}^{AO}_{-i}}\Big[ Q^{\pi_{i},\bm{\pi}^{AO}_{-i},  \bm{\beta}^{O}, \bm{\alpha};\mu_{i}}_{i|t,t'}(h_{i,t}[\hat{a}^{(t)}_{i},\bm{a}^{t}_{-i}]|\theta_{i})\Big]\Bigg)
    +\sum_{i\in \mathcal{N},\beta_{i}} \Xi^{\beta_{i}}_{i;t,t'}\Bigg(\zeta_{t,t'}^{ \bm{\alpha}^{OIL}}[\beta_{i}]- J_{i}^{\bm{\pi}^{AO},\bm{\beta}^{O}, \bm{\alpha}}(g_{0}|\theta_{i})\\
    +&\mathbb{E}^{\bm{\alpha}^{\text{OIL}}} \Big[
        V^{\bm{\pi}^{AO}, \beta_{i}, \bm{\beta}^{O}_{-i}, \bm{\alpha};\mu_{i}}_{i|t,t'}(\ell_{i,t}[\hat{\omega}^{(t)}_{i}, \omega^{k;(t)}_{i}]|\theta_{i})\Big]\Bigg), 
    \end{split}
\end{equation}
%

Due to the Lagrangian sufficiency theorem (see, e.g., \cite{boyd2004convex}), one way to design an OIL-P signaling rule $\bm{\alpha}$ is to make sure that there exist a pair $\bm{\Lambda}_{t,t'}$ and $\bm{\Xi}_{t,t'}$, such that 
$$
\begin{aligned}
&\min_{\bm{\pi}^{*},  \bm{\beta}^{*},  \bm{V}^{*}} L^{\bm{\beta}^{*}, \bm{\alpha} }_{\bm{\pi}^{*} }(\bm{\Lambda}_{t,t'}, \bm{\Xi}_{t,t'}, \bm{V}^{*}; \bm{\delta}_{t,t'}^{\bm{\alpha} }, \bm{\zeta}_{t,t'}^{\bm{\alpha} }|\bm{\theta}) 
= L^{\bm{\beta}^{O}, \bm{\alpha} }_{\bm{\pi}^{AO}}(\bm{\Lambda}_{t,t'}, \bm{\Xi}_{t,t'},\bm{V}^{\bm{\pi}^{AO}, \bm{\beta}^{O}, \bm{\alpha}}; \bm{\delta}_{t,t'}^{\bm{\alpha} }, \bm{\zeta}_{t,t'}^{\bm{\alpha} } |\bm{\theta}).
\end{aligned}
$$
The Lagrangian sufficiency theorem states that, with such $\bm{\alpha}$, the admissible O-PBME is also an optimal solution of the problem in Theorem \ref{thm:constrained_IDP}. 
From the definition of admissible O-PBME, it is straightforward to see that $L^{\bm{\beta}^{O}, \bm{\alpha} }_{\bm{\pi}^{AO}}(\bm{\Lambda}_{t,t'}, \bm{\Xi}_{t,t'},\bm{V}^{\bm{\pi}^{AO}, \bm{\beta}^{O}, \bm{\alpha}}; \bm{\delta}_{t,t'}^{\bm{\alpha} }, \bm{\zeta}_{t,t'}^{\bm{\alpha}}$  $|\bm{\theta})=0$; i.e., the slack variables can be written as: for any $t,t'\geq 0$,
\begin{equation}\label{eq:slack_pi}
    \begin{aligned}
&\delta_{\bm{\pi}^{AO};t,t'}^{\bm{\beta}^{O}, \bm{\alpha} }[\pi_{i}] = V_{i}^{\bm{\pi}^{AO},\bm{\beta}^{O}, \bm{\alpha}}(g_{0},\omega^{k}_{i,0}|\theta_{i})-\mathbb{E}^{\mu_{i}}_{\bm{\pi}^{AO}_{-i}}\Big[ Q^{\pi_{i},\bm{\pi}^{AO}_{-i},  \bm{\beta}^{O}, \bm{\alpha};\mu_{i}}_{i|t,t'}(h_{i,t}[\hat{a}^{(t)}_{i},\bm{a}^{t}_{-i}]|\theta_{i})\Big]\Big|_{\bm{\Lambda}_{t,t'}, \bm{\Xi}_{t,t'}},
\end{aligned}
\end{equation}
\begin{equation}\label{eq:slack_beta}
    \begin{aligned}
&\zeta_{\bm{\pi}^{AO};t,t'}^{\bm{\beta}^{O}, \bm{\alpha} }[\beta_{i}]= J_{i}^{\bm{\pi}^{AO},\bm{\beta}^{O}, \bm{\alpha}}(g_{0}|\theta_{i})-\mathbb{E}^{\bm{\alpha}^{\text{OIL}}} \Big[
        V^{\bm{\pi}^{AO}, \beta_{i}, \bm{\beta}^{O}_{-i}, \bm{\alpha};\mu_{i}}_{i|t,t'}(\ell_{i,t}[\hat{\omega}^{(t)}_{i}, \omega^{k;(t)}_{i}]|\theta_{i})\Big]\Big|_{\bm{\Lambda}_{t,t'}, \bm{\Xi}_{t,t'}}.
\end{aligned}
\end{equation}
Here, we add the profile $<\bm{\beta}^{O}, \bm{\pi}^{AO}>$ in the superscript and the subscript of the slack variables to show their dependence on $<\bm{\beta}^{O}, \bm{\pi}^{AO}>$.
Hence, $<\bm{\beta}^{O}, \bm{\pi}^{AO}, \bm{V}^{\bm{\pi}^{AO}, \bm{\beta}^{O}, \bm{\alpha} }>$ and $\bm{\delta}^{\bm{\alpha} }$ in (\ref{eq:slack_pi}) and $\bm{\zeta}^{\bm{\alpha} }$ in (\ref{eq:slack_beta}) make the constraints (\ref{eq:sequential_constraint_1}) and (\ref{eq:sequential_constraint_2}) binding.

Define, for any $\bm{\beta}$, $\bm{\pi}$, $g\in\mathcal{G}$, $\bm{a}\in\bm{\mathcal{A}}$, $\bm{\omega}\in \Omega^{n}$, $\bm{\omega}^{k}\in \Omega^{n}$, $\bm{\theta}\in \bm{\Theta}$,
\begin{equation}\label{eq:occupancy_reward}
    \begin{aligned}
    U^{\bm{\beta}}_{\bm{\pi}}(g, \bm{a}, \bm{\omega}|\bm{\omega}^{k}, \bm{\theta}) 
    \equiv \sum_{i\in\mathcal{N}} R_{i}(g,\bm{a}, \omega_{i}|\theta_{i})\bm{\rho}^{\bm{\beta}}_{\bm{\pi}}(g, \bm{a}, \bm{\omega}|\bm{\omega}^{k}, \bm{\theta}),
    \end{aligned}
\end{equation}
with $U^{\bm{\beta}}_{\bm{\pi}}(g, \bm{a},  \bm{\omega}^{k}| \bm{\theta}) = U^{\bm{\beta}}_{\bm{\pi}}(g, \bm{a}, \bm{\omega}^{k}| \bm{\omega}^{k}, \bm{\theta})$,
where $\bm{\rho}^{\bm{\beta}}_{\bm{\pi}}(\cdot|\bm{\omega}^{k}, \bm{\theta})$ is given in (\ref{eq:short_occupancy}).

Based on the result of Lemma \ref{lemma:occupancy_relationship}, we have the following proposition.

\begin{proposition}\label{prop:solution_dual_variable}
Let $\bm{\Lambda}^{*}_{t,t'}$ and $\bm{\Xi}^{*}_{t,t'}$ denote the dual variables such that $L^{\bm{\beta}^{O}, \bm{\alpha} }_{\bm{\pi}^{AO}}(\bm{\Lambda}_{t,t'}, \bm{\Xi}_{t,t'},\bm{V}^{\bm{\pi}^{AO}, \bm{\beta}^{O}, \bm{\alpha}};$ $ \bm{\delta}_{\bm{\pi}^{AO};t,t'}^{\bm{\beta}^{O}, \bm{\alpha} },$ $ \bm{\zeta}_{\bm{\pi}^{AO};t,t'}^{\bm{\beta}^{O}, \bm{\alpha} }$  $|\bm{\theta})=0$ for $\delta_{\bm{\pi}^{AO};t,t'}^{\bm{\beta}^{O}, \bm{\alpha} }[\pi_{i}]$ and $\zeta_{\bm{\pi}^{AO};t,t'}^{\bm{\beta}^{O}, \bm{\alpha} }[\beta_{i}]$ given in (\ref{eq:slack_pi}) and (\ref{eq:slack_beta}). The followings hold: for any $\pi_{i}$, $\beta_{i}$, $i\in\mathcal{N}$,
\begin{equation}\label{eq:slack_pi_limit}
    \begin{aligned}
    &\lim_{t-t'\rightarrow \infty }\delta_{\bm{\pi}^{AO};t,t'}^{\bm{\beta}^{O}, \bm{\alpha} }[\pi_{i}]= \sum_{\substack{g,\bm{a}, \bm{\omega}^{k},\\
    i\in\mathcal{N}}}\bm{\alpha}(\bm{\omega}^{k}|g,\bm{\theta})U^{\bm{\beta}^{O}}_{\bm{\pi}^{AO}}(g, \bm{a},\bm{\omega}^{k}| \bm{\theta})-\sum_{\substack{g,\bm{a}, \bm{\omega}^{k},\\
    i\in\mathcal{N}}}\bm{\alpha}(\bm{\omega}^{k}|g,\bm{\theta})
    U^{\bm{\beta}^{O}}_{\pi_{i}, \bm{\pi}^{AO}_{-i}}(g,a_{i},\bm{a}_{-i}, \bm{\omega}^{k}| \bm{\theta}),
    \end{aligned}
\end{equation}
\begin{equation}\label{eq:slack_beta_limit}
    \begin{aligned}
    \lim_{t-t' \rightarrow \infty} \zeta_{\bm{\pi}^{AO};t,t'}^{\bm{\beta}^{O}, \bm{\alpha} }[\beta_{i}] &= \sum_{\substack{g,\bm{a}, \bm{\omega}^{k},\\
    i\in\mathcal{N}}}\bm{\alpha}(\bm{\omega}^{k}|g,\bm{\theta})U^{\bm{\beta}^{O}}_{\bm{\pi}^{AO}}(g, \bm{a},\bm{\omega}^{k}| \bm{\theta})\\
    &-\sum_{\substack{g,\bm{a}, \bm{\omega}^{k},\\
    i\in\mathcal{N}}}\bm{\alpha}(\bm{\omega}^{k}|g,\bm{\theta}) U^{\beta_{i},\bm{\beta}^{O}_{-i}}_{ \bm{\pi}^{AO} }(g,\bm{a}, \beta_{i}(g, \omega^{k}_{i}, \theta_{i}),\bm{\omega}^{k}_{-i}| \bm{\omega}^{k}, \bm{\theta}),
    \end{aligned}
\end{equation}
and the corresponding dual variables are $\lim\limits_{t-t'\rightarrow \infty} \Lambda^{\pi_{i};*}_{i;t,t'} = \lambda^{\bm{\alpha},\bm{\beta}^{O};\mu_{i}}_{[i], \pi_{i},\bm{\pi}^{AO}_{-i}, \infty}(\cdot|\theta_{i})$ and   $\lim\limits_{t-t'\rightarrow \infty}$ $\Xi^{\beta_{i};*}_{i;t,t'}=$ $ \bar{\lambda}^{\bm{\alpha},\beta_{i},\bm{\beta}^{O}_{-i};\mu_{i}}_{[i], \bm{\pi}^{AO}, \infty}(\cdot|\theta_{i})$.
\end{proposition}
%
%


Proposition \ref{prop:solution_dual_variable} describes an asymptotic situation in which the length of sequence $h_{i,t}$ of $Q^{\bm{\pi}, \bm{\beta}, \bm{\alpha};\mu_{i}}_{i|t,t'}$ is much larger than the length of sequence $\ell'_{i,t}$ of $V^{\bm{\pi}, \bm{\beta}, \bm{\alpha};\mu_{i}}_{i|t,t'}$ in the extended Bellman equations (\ref{eq:sequential_occupancy_Q})-(\ref{eq:sequential_occupancy_V}).
We remove the subscripts $t$ and $t'$ if $t-t'\rightarrow \infty$ in the notations of the slack and the dual variables.
This asymptotic result motivates a design regime for the signaling rule:
\begin{theorem}\label{thm:strong_OIL_P}
%
%
%
Define a set of signaling rules, for any $\bm{\kappa}$,
\begin{equation}\label{eq:set_of_OIL_signaling_rule_1} 
    \begin{aligned}
    &\bm{A}^{O}[\bm{\kappa}]\equiv \Big\{\bm{\alpha}^{*}: \forall \bm{\pi}^{*}\in \bm{\Pi}^{O},\bm{\alpha}^{*}\in\arg\max_{\bm{\alpha}} \min_{\bm{\pi},\bm{\beta}} \sum_{\pi,\beta}\big(\bm{\delta}^{\bm{\beta}^{O}, \bm{\alpha} }_{\bm{\pi}^{*}}[\bm{\pi}]+ \bm{\zeta}^{\bm{\beta}^{O}, \bm{\alpha} }_{\bm{\pi}^{*}}[\bm{\beta}]\big)\text{ and } \\
    & \sum_{\bm{\omega}^{k}}\bm{\rho}^{\bm{\beta}^{O}}_{\bm{\pi}^{*}}(g,\bm{a}, \bm{\omega}^{k}|\bm{\theta})\bm{\alpha}^{*}(\bm{\omega}^{k}|g,\bm{\theta}) =\bm{\rho}^{\bm{\kappa}}(g,\bm{a} |\bm{\theta})\Big\}.
    \end{aligned}
\end{equation}
Given a goal $\bm{\kappa}$, a signaling rule $\bm{\alpha}^{*}\in \bm{A}^{O}[\bm{\kappa}]$ that induces $<\bm{\beta}^{O}, \bm{\pi}^{*}>$ is OIL-P if $L^{\bm{\beta}^{O}, \bm{\alpha}^{*} }_{\bm{\pi}^{*}}(\bm{\Lambda}, $ $ \bm{\Xi},\bm{V}^{\bm{\pi}^{*}, \bm{\beta}^{O}, \bm{\alpha}^{*}};$  $ \bm{\delta}_{\bm{\pi}^{O}}^{\bm{\beta}^{O},\bm{\alpha}^{*}}, \bm{\zeta}^{\bm{\beta}^{O},\bm{\alpha}^{*}}_{\bm{\pi}^{*}}$  $|\bm{\theta})=0$.
\end{theorem}


In Theorem \ref{thm:strong_OIL_P}, the set $\bm{A}^{O}_{t,t'}[\bm{\kappa}]$ in (\ref{eq:set_of_OIL_signaling_rule_1}) characterizes a design regime to determine signaling rules that realize the principal's goal $\bm{\kappa}$ while each agent $i$ has incentive to use the obedient selection rule for every possible observation $o_{i}$.
The $\bm{\alpha}^{*}$'s maximizing the minimum of the slack variables (\ref{eq:slack_pi}) and (\ref{eq:slack_beta}) makes any of its induced $<\bm{\beta}^{O}, \bm{\pi}^{AO}>$ be a feasible solution of the problem (\ref{eq:objective_IDP_V1})-(\ref{eq:idp_admissible_constraint}).
From the Lagrangian sufficiency theorem, the condition $L^{\bm{\beta}^{*}, \bm{\alpha}^{*} }_{\bm{\pi}^{*}}(\bm{\Lambda}_{t,t'}, $ $ \bm{\Xi}_{t,t'},\bm{V}^{\bm{\pi}^{*}, \bm{\beta}^{O}, \bm{\alpha}^{*}}; \bm{\delta}^{\bm{\alpha}^{*}}, \bm{\zeta}^{\bm{\alpha}^{*}}$  $|\bm{\theta})=0$ yields that $<\bm{\beta}^{O}, \bm{\pi}^{*}>$ and the corresponding $\bm{V}^{\bm{\pi}^{*}, \bm{\beta}^{O}, \bm{\alpha}^{*}}$ (that satisfies (\ref{eq:bellman_V_1}) (\ref{eq:bellman_Q_1})) is a solution of (\ref{eq:objective_IDP_V1})-(\ref{eq:idp_admissible_constraint}) with $Z^{\text{OIL-P}}(\bm{\pi}^{AO}, \bm{\beta}^{O},$ $\bm{V}^{\bm{\alpha}^{\text{OIL}}}; \bm{\alpha}^{\text{OIL}}, \bm{\kappa}) = 0$.
%





\section{Optimal Information Design}\label{sec:optimal_information_design}

In this section, we introduce the optimality criterion of the principal's goal $\bm{\kappa}$ and define the optimal information design problem for the principal in a Markov game $\bm{M}[\bm{\alpha}]$. 
We define the one-stage payoff function of the principal is $u^{k}:\bm{\mathcal{A}}\times \mathcal{G}\times \bm{\Theta}\mapsto \mathbb{R}$, such that $u^{k}(\bm{a}, g; \bm{\theta})$ gives the immediate payoff for the principal when the state is $g$ and the agents with type $\bm{\theta}$ take actions $\bm{a}$.
The principal's goal $\bm{\kappa}$ is the probability distribution of the agents' joint actions in the equilibrium conditioned only on the global state and the agents' types.
Hence, the information structure that matters for the principal's goal selection problem is given as $\mathcal{O}^{k}\equiv <\mathcal{G}, \bm{\Theta}, \mathcal{T}_{g}, d_{g}, d_{\theta}>$.
The principal chooses a goal by maximizing her expected payoff ($\hat{\gamma}$-discounted, $\hat{\gamma}\in(0,1]$), given as:
\begin{equation}\label{eq:principal_optimal_info_design_v1}
    \begin{aligned}
    Z^{k}(\bm{\kappa}| \mathcal{O}^{k})\equiv \mathbb{E}^{\bm{\kappa}}\Big[\sum_{t=0}^{\infty}\hat{\gamma}^{t} u^{k}(\bm{a}_{t}, g_{t}, \bm{\theta})\Big| \mathcal{O}^{k}\Big].
    \end{aligned}
\end{equation}

The principal's goal $\bm{\kappa}$ is chosen such that her expected payoff (\ref{eq:principal_optimal_info_design_v1}) is maximized. 
However, the principal cannot force the agents to take the actions or directly program agents' actions according to $\bm{\kappa}$; instead, she uses information design to elicit the agents to take actions that coincide with her goal $\bm{\kappa}$.
The following theorem discovers an important relationship between the goal $\bm{\kappa}$ and the agents' equilibrium.

\begin{theorem}\label{thm:revelation_equivalence}
The profile $<\bm{\beta}^{O}, \bm{\pi}^{AO}>$ is an admissible O-PBME if and only if the goal $\bm{\kappa}$ is a BMCE (Definition \ref{def:BMCE}) and there exists an $\bm{\alpha}^{*}\in \bm{A}^{O}[\bm{\kappa}]$ such that $L^{\bm{\beta}^{O}, \bm{\alpha}^{*} }_{\bm{\pi}^{*}}(\bm{\Lambda}, $ $ \bm{\Xi},\bm{V}^{\bm{\pi}^{*}, \bm{\beta}^{O}, \bm{\alpha}^{*}};$  $ \bm{\delta}_{\bm{\pi}^{O}}^{\bm{\beta}^{O},\bm{\alpha}^{*}}, \bm{\zeta}^{\bm{\beta}^{O},\bm{\alpha}^{*}}_{\bm{\pi}^{*}}$  $|\bm{\theta})=0$.
\end{theorem}
%


Theorem \ref{thm:revelation_equivalence} strictly generalizes the Bergemann and Morris's characterization of Bayes' correlated equilibrium for incomplete-information static game (Theorem 1 of \cite{bergemann2016bayes}) to our Markovian settings where agents select signals and take actions.
Basically, the BMCE characterizes all the possible O-PBME that could arise under all signaling rules in $\bm{A}^{O}[\bm{\kappa}]$ in which $\bm{\kappa}$ is any BMCE.
Hence, the principal's goal selection problem is a BMCE selection problem.


Suppose the principal's $\bm{\alpha}$ induces a profile $<\bm{\beta}, \bm{\pi}>$. With a slight abuse of notation, we define the principal's expected payoff from the agents' behaviors by $<\bm{\beta}, \bm{\pi}>$ as follows:
\begin{equation}\label{eq:transformed_principal}
    \begin{aligned}
    &Z^{k}(\bm{\alpha}, \bm{\beta},\bm{\pi}| \bm{\mathcal{O}}^{\bm{\alpha}})\equiv \mathbb{E}\Big[\sum^{\infty}_{t=0}\hat{\gamma}^{t}u^{k}(\bm{a}_{t}, g_{t};\bm{\theta})\bm{\pi}(\bm{a}_{t}|g_{t}, \bm{\beta}(g_{t},\bm{\omega}^{k}_{t}, \bm{\theta}), \bm{\theta})\Big| \bm{\mathcal{O}}^{\bm{\alpha}}\Big],
    \end{aligned}
\end{equation}
with $Z^{k}(\bm{\alpha}, \bm{\pi}| \bm{\mathcal{O}}^{\bm{\alpha}})=Z^{k}(\bm{\alpha}, \bm{\beta}^{O},\bm{\pi}| \bm{\mathcal{O}}^{\bm{\alpha}})$,
where the information structure is $\bm{\mathcal{O}}^{\bm{\alpha}}=<\mathcal{G}, \Omega^{m}, \bm{\Theta},$  $\mathcal{T}_{g}, \mathcal{P}^{-k},$ $ \bm{\alpha}, d_{g}, d_{\theta}>$.
We refer to (\ref{eq:transformed_principal}) as the principal's \textit{transformed problem}.
Define a set of PBME profiles $\mathtt{PBME}[\bm{\alpha}]$, when the signaling rule is $\bm{\alpha}$:
\begin{equation}\label{eq:set_PBME_UK}
    \begin{aligned}
    \mathtt{PBME}&[\bm{\alpha}]\equiv \Big\{\bm{\beta}^{*}, \bm{\pi}^{*}: \forall \beta_{i},\pi_{i}, i\in \mathcal{N}Z^{k}(\bm{\alpha}, \bm{\beta}^{*}, \bm{\pi}^{*}|\bm{\mathcal{O}}^{\bm{\alpha}})\geq  Z^{k}(\bm{\alpha}, \beta_{i}, \bm{\beta}^{*}_{-i}, \pi_{-i},\bm{\pi}^{*}|\bm{\mathcal{O}}^{\bm{\alpha}}) \Big\}.
    \end{aligned}
\end{equation}
We will write $\mathtt{PBME}[\bm{\alpha}, \bm{\beta}^{*}]$ as a set of PMBE policy profile when agents use equilibrium selection rule profile $\bm{\beta}^{*}$ (i.e., $\bm{\beta}^{*}$ is used in both sides of the inequality in (\ref{eq:set_PBME_UK})), given $\bm{\alpha}$.

We define a set of signaling rules, for any policy profile $\bm{\pi}^{*}$,
\begin{equation}\label{eq:info_design_optimization_v2}
    \begin{aligned}
    &\bm{A}^{O}[\bm{\pi}^{*}]\equiv \Big\{\bm{\alpha}^{*}: \bm{\alpha}^{*}\in\arg\limits_{\bm{\alpha}}\max \min_{\bm{\pi},\bm{\beta}} \sum_{\pi,\beta}\big(\bm{\delta}^{\bm{\beta}^{O}, \bm{\alpha} }_{\bm{\pi}^{*}}[\bm{\pi}]+ \bm{\zeta}^{\bm{\beta}^{O}, \bm{\alpha} }_{\bm{\pi}^{*}}[\bm{\beta}]\big)
    \Big\}.
\end{aligned}
\end{equation}
Hence, $\bm{A}^{O}[\bm{\pi}^{*}]$ is the set $\bm{A}^{O}[\bm{\kappa}]$ in (\ref{eq:set_of_OIL_signaling_rule_1}) without the admissibility condition.

Theorem \ref{thm:revelation_equivalence} motivates the following transformation of the principal's $\mathtt{BMCE}$ selection problem to an information design problem:
\begin{equation}\label{eq:principal_equivalence}
    \begin{aligned}
    \max_{\bm{\kappa}\in \mathtt{BMCE}} Z^{k}(\bm{\kappa}|\mathcal{O}^{k}) = \max_{\bm{\alpha}\in \bm{A}^{O}[\bm{\pi}]}\max_{\bm{\pi}\in  \mathtt{PBME}[\bm{\alpha}, \bm{\beta}^{O}]}Z^{k}(\bm{\alpha}, \bm{\pi}|\bm{\mathcal{O}}^{\bm{\alpha}}).
    \end{aligned}
\end{equation}

Suppose $\bm{\kappa}^{*}$ is an optimal BCE for the principal.
Suppose additionally that $\bm{\alpha}^{*}$ is a solution to the RHS of (\ref{eq:principal_equivalence}). 
However, choosing a signaling rule from $\bm{A}^{O}[\bm{\kappa}]$ in (\ref{eq:set_of_OIL_signaling_rule_1}) is in general a sufficient condition for OIL-P.
Hence, for any $<\bm{\beta}^{*}, \bm{\pi}^{*}>\in \mathtt{PBME}[\bm{\alpha}^{*}]$, we may have $L^{\bm{\beta}^{*}, \bm{\alpha}^{*} }_{\bm{\pi}^{*}}(\bm{\Lambda}, $ $ \bm{\Xi},\bm{V}^{\bm{\pi}^{*}, \bm{\beta}^{*}, \bm{\alpha}^{*}};$  $ \bm{\delta}_{\bm{\pi}^{*}}^{\bm{\beta}^{*},\bm{\alpha}^{*}}, \bm{\zeta}^{\bm{\beta}^{*},\bm{\alpha}^{*}}_{\bm{\pi}^{*}}$  $|\bm{\theta})\neq0$.
That is, the principal's optimal goal may not be realized by the designed signaling rule $\bm{\alpha}^{*}$, i.e.,
$
Z^{k}(\bm{\kappa}^{*}|\mathcal{O}^{k}) \geq Z^{k}(\bm{\alpha}^{*},  \bm{\beta}^{*},\bm{\pi}^{*}|\bm{\mathcal{O}}^{\bm{\alpha}^{*}}).
$
When the choice of the signaling rule characterized in $\bm{A}^{O}[\bm{\kappa}]$ in (\ref{eq:set_of_OIL_signaling_rule_1}) cannot guarantee the OIL-P, we consider a notion of $\epsilon$-OIL-P which is a relaxation of OIL-P.

\begin{definition}[$\epsilon$-OIL-P]
We say that a signaling rule $\bm{\alpha}_{\epsilon}$ is $\epsilon$-OIL-P if it induces a profile $<\bm{\beta}^{*}_{\epsilon}, \bm{\pi}^{*}_{\epsilon}>$ with $\mu_{i}$,
%
%
such that the belief $\mu_{i}$ is updated according to (\ref{eq:belief}) and, for all $\pi_{i}$, $\beta_{i}$, $i\in\mathcal{N}$,
\begin{equation}
    \begin{aligned}
    Z^{k}(\bm{\alpha}_{\epsilon}, \bm{\beta}^{*}_{\epsilon}, \bm{\pi}^{*}_{\epsilon})+\epsilon \geq Z^{k}(\bm{\alpha}_{\epsilon}, \beta_{i}, \bm{\beta}^{*}_{-i;\epsilon},\pi_{i}, \bm{\pi}^{*}_{-i;\epsilon}).
    \end{aligned}
\end{equation}
\end{definition}

Let $v^{k}(\cdot; \bm{\beta}, \bm{\pi}, \bm{\alpha}): \mathcal{G}\times \bm{\Theta}\mapsto \mathbb{R}$ denote the state-value function associated with the principal's transformed problem (\ref{eq:transformed_principal}).
Define, for any $g\in\mathcal{G}$, $\bm{\theta}\in \bm{\Theta}$,
\begin{equation}
    \begin{aligned}
\Psi(g, \bm{\theta};\bm{\pi}, \bm{\beta}, \bm{\alpha})\equiv& v^{k}(g, \bm{\theta}; \bm{\beta}, \bm{\pi}, \bm{\alpha})
-\mathbb{E}^{\mu_{i}}_{\bm{\pi}}\Big[ \sum_{i} R_{i}(g, \bm{a}, \beta_{i}(g,\omega^{k},\theta_{i}), \theta_{i}) 
    + \gamma \sum_{g'} \mathcal{T}_{g}(g'|g, \bm{a} ) v^{k}(g',\bm{\theta};\bm{\beta}, \bm{\pi}, \bm{\alpha})   \Big| \bm{\mathcal{O}}^{\bm{\alpha}}\Big].
\end{aligned}
\end{equation}
If $\Psi(g, \bm{\theta};\bm{\pi}, \bm{\beta}, \bm{\alpha}) = 0$, the state-value function $v^{k}(\cdot; \bm{\beta}, \bm{\pi}, \bm{\alpha})$ is the optimal state-value function (i.e., solution to the Bellman optimality equation).
%
%

\begin{proposition}\label{prop:epsilon_OIL-P}
Suppose the signaling rule $\bm{\alpha}$ induces $<\bm{\beta}^{O}_{\epsilon}, \bm{\pi}^{AP}_{\epsilon}>$ that is feasible with respect to the constraints (\ref{eq:IPD_constraint_1_1}) and (\ref{eq:IPD_constraint_1_2}).
Let
$$\Phi(\bm{\theta}; \bm{\pi}, \bm{\beta}, \bm{\alpha}) \equiv \mathbb{E}\Big[\Psi(g, \bm{\theta};\bm{\pi}, \bm{\beta}, \bm{\alpha})\Big|\bm{\mathcal{O}}^{\bm{\alpha}}\Big].
$$
Then, $<\bm{\beta}^{O}_{\epsilon}, \bm{\pi}^{AP}_{\epsilon}>$ forms an $\epsilon$-admissible O-PBME, with $\epsilon= \frac{\Phi(\bm{\theta}; \bm{\pi}, \bm{\beta}, \bm{\alpha}) }{1-\hat{\gamma}}$. The corresponding signaling rule $\bm{\alpha}$ is $\epsilon$-OIL-P.
\end{proposition}


Proposition \ref{prop:epsilon_OIL-P} characterizes the $\epsilon$-OIL-P as an approximation to the optimal information design given in Theorem \ref{thm:strong_OIL_P}. 
Decent algorithms used to solve the information design problem may lead to approximated equilibrium behaviors and Proposition \ref{prop:epsilon_OIL-P} implies that when $\Phi(\bm{\theta}; \bm{\pi}, \bm{\beta}, \bm{\alpha})$ is small enough relative to $1-\hat{\gamma}$, the computational outcome from the algorithm is good enough (i.e., $\epsilon\rightarrow 0$).

\section{Conclusion}\label{sec:conclusion}

This work is the first to propose an information design principle for incomplete-information dynamic games in which each agent makes coupled decisions of selecting a signal and taking an action at each period of time.
We have formally defined a novel information design problem for the indirect and the direct settings and have restricted attention to the direct one due to the obedient principle.
The notion of obedient implementability has been introduced to capture the optimality of the direct information design in the equilibrium concept of obedient perfect Bayesian Markov Nash equilibria (O-PBME).
By characterizing the obedient implementability, we have proposed an approach to determining the information structure by maximizing the optimal slack variables from the optimality of the agents' equilibrium behaviors.
Our representation result formulates the principal's optimal Bayesian Markov correlated equilibrium selection in terms of information design implementable in O-PBME.

\bibliographystyle{unsrt}  
\bibliography{references}  

\end{document}